\documentclass[aps,prd,twocolumn,nofootinbib,superscriptaddress,floatfix]{revtex4-2}

\usepackage{amsmath,amssymb,amsfonts}
\usepackage{bm}
\usepackage{graphicx}
\usepackage{multirow}
\usepackage[colorlinks=true,linkcolor=blue,citecolor=blue,urlcolor=blue]{hyperref}

\newcommand{\K}{\mathrm{K}}
\newcommand{\dd}{\mathrm{d}}
\newcommand{\calP}{\mathcal P}
\newcommand{\calQ}{\mathcal Q}
\newcommand{\calW}{\mathcal W}
\newcommand{\calH}{\mathcal H}
\newcommand{\calR}{\mathcal R}

\newcommand{\epsD}{\epsilon_{\Delta}}

\begin{document}
	
	\title{Deviations from Kerr: Polar critical curves and photon rings in a class of separable spacetimes}

	\author{Hao-Peng Yan} \email{Corresponding author: yanhaopeng@tyut.edu.cn}
	\author{Xiang-Qian Li} 
	\author{Xiao-Jun Yue} %\email{yuexiaojun@tyut.edu.cn} 
	\affiliation{College of Physics and Optoelectronic Engineering, Taiyuan University of Technology, Taiyuan 030024, China}
	
%	\date{\today}
	
	\begin{abstract}
Black-hole critical curves and photon rings provide complementary probes of
strong-field geometry, motivating a unified analytic description beyond Kerr
and direct comparison with Kerr. We develop such a description for polar
observers in a broad class of stationary, axisymmetric spacetimes with
separable null geodesics. We organize the metric dependence in two
complementary ways: through optical combinations naturally selected by the
null dynamics and through a Kerr-relative representation suited to weak
deviations from Kerr. This yields exact expressions for the polar critical
radius, Lyapunov exponent, time delay, and rotation parameter, together with
general first-order formulas applicable once the deformation functions of a
specific metric are specified. At first order, the combined polar observables
depend only on a finite set of local values and radial derivatives of these
functions at the critical orbit, rather than on their full radial profiles.
\end{abstract}

\maketitle

\section{Introduction}
\label{sec:introduction}

Black-hole photon rings are produced by photons that remain close to unstable bound null geodesics before reaching a distant observer. The critical curve is the limiting locus of the high-order photon subrings. For a Kerr black hole, Hamilton--Jacobi separability permits an analytic description of the photon shell, the critical curve, and the asymptotic high-order ring structure \cite{Carter1968,Bardeen1973,Teo2003,PerlickTsupko2022,GrallaLupsasca2020}. Successive photon subrings encode geometric information through their demagnification, relative rotation, and time delay \cite{GrallaLupsasca2020,Johnson2020,HadarJohnsonLupsascaWong2021}.

The Event Horizon Telescope observations of M87* and Sgr~A* have renewed interest in determining which aspects of the strong-field geometry can be constrained by image-domain measurements \cite{EHTM87,EHTSgrA}. The reconstructed emission ring is not identical to the geometric critical curve or to an individual high-order photon subring: plasma emission, radiative transfer, scattering, and finite angular resolution all enter the observed brightness distribution \cite{GrallaHolzWald2019,Johnson2020,ChenJingQianWang2023}. Dedicated null-hypothesis and horizon-scale analyses have translated image-size measurements into constraints on departures from Kerr \cite{PsaltisOzelChanMarrone2015,EHTM87Metric2020,EHTSgrAMetric2022}. The use of the photon-ring shape as a precision test of the Kerr
hypothesis has been developed in Refs.~\cite{GrallaLupsascaObservable2020,
	GrallaLupsascaMarrone2020,PaugnatLupsascaVincentWielgus2022}, including its dependence on
source structure and viewing geometry. 

From the perspective of spacetime modeling, rotating non-Kerr geometries
have been studied one metric at a time through their shadows, critical
photon structures, and related optical phenomena  
\cite{JohannsenPsaltisImages2010,
	CunhaHerdeiroRaduRunarsson2015,
	YounsiZhidenkoRezzollaKonoplyaMizuno2016,
	MedeirosPsaltisOzel2020,
	GuoObersYan2018,WangXuWei2019,WeiLiu2021,
	LongWangChenJing2019,MengKuangTang2022,
	LiMirzaevAbdujabbarovMalafarinaAhmedovHan2022,
	WangChenJing2021,
	StaelensMayersonBacchiniRipperdaKuchler2023,HeLiYangZeng2025,
	LongChenJing2026,WeiWangZhangLiuMann2026}.
Strong-field ring structures in modified spacetimes have also been explored
from holographic wave-optics perspectives
\cite{Gui:2024owz,Zeng:2024ptv}. 
Detailed numerical ray-tracing and radiative-transfer calculations have
further explored how departures from Kerr appear in full images, inner
shadows, higher-order structures, and polarization observables for specific
modified or parametrized spacetimes
\cite{BroderickJohannsenLoebPsaltis2014,
	MizunoYounsiFrommEtAl2018,HouZhangYanGuoChen2022,
	SunLiuQianChenYue2023,ZhangHouGuoChen2024,WangWangZhangGuo2024,ZhengWuLiJiang2025,
	LiZhengHeJiang2025,HeYangZengChang2026,
	YangYeZeng2026,
	GanWangWuYang2021,Huang:2024bar,HeYeZengLiXu2025,ZengYangAslamSaleem2026,WangWangWangGuo2026}.
Such calculations are valuable for displaying the phenomenology of
particular theories or effective models, but a model-by-model comparison
does not by itself reveal which combinations of the metric control a given
observable, which deformation directions are invisible, or which non-Kerr effects can
be absorbed into changes of the Kerr parameters. Direct comparisons of
non-Kerr photon-ring shapes and Lyapunov exponents have also shown that
the instability exponent can retain appreciable sensitivity even when the
low-inclination ring shape remains close to Kerr
\cite{StaelensMayersonBacchiniRipperdaKuchler2023}.
A metric-function-level analytic description can retain the map from
spacetime structure to observables before the result is compressed into
a small number of model parameters.

Several broad metric parametrizations provide a natural starting point for such a program. Generic rotating parametrizations and symmetry-preserving constructions have been developed with different levels of restriction on separability and hidden symmetries \cite{CardosoPaniRico2014,KonoplyaRezzollaZhidenko2016,PapadopoulosKokkotas2018,CarsonYagi2020}. Within the separable class, the Johannsen metric preserves a Carter-like constant while encoding deviations from Kerr in freely specifiable deformation functions \cite{Johannsen2015,JohannsenPhotonRings2013}. Its null-geodesic dynamics, shadow morphology, EHT constraints, and ray-traced disk images have subsequently been investigated for the standard leading radial deformation parameters \cite{MengFanLiHanZhang2023}. Salehi, Broderick, and Georgiev (SBG) recast a radial sector of this separable structure in terms of four nonparametric functions $N(r)$, $F(r)$, $B(r)$, and $f_{\rm SBG}(r)$ (denoted $f$ in their original notation) with direct geometric interpretations, and identified the combinations sampled by polar shadow and photon-ring observables \cite{SalehiBroderickGeorgiev2024}. General-inclination photon-ring critical parameters were subsequently developed in the same framework \cite{SalehiWaliaChangKocherlakota2025,WaliaKocherlakotaChangSalehi2025}. The Kerr-off-shell (KOS) construction preserves the full Kerr
hidden-symmetry tower and Petrov type-D structure, encoding deviations in one
radial and one polar function; it yields an exact parametrization of the critical curve at general inclination and exhibits strong
mass--spin--deformation degeneracies in circlipse fits
\cite{AchourGourgoulhonRoussille2025,
	GrallaLupsascaObservable2020,GrallaLupsascaMarrone2020}. 
More recently, Kobialko and Gal'tsov developed an analytical perturbation theory for shadow morphology in general separable Benenti-type metrics, expanding about Schwarzschild in both rotation and non-Kerr deformations and including separable plasma effects \cite{KobialkoGaltsov2026}. In parallel, numerical work has developed efficient ray-tracing and
parameter-inference frameworks for broad classes of stationary non-Kerr
geometries, providing a route from parametrized spacetimes to horizon-scale
observational constraints and forecasts for future imaging experiments
\cite{LiuLiuWuLiu2026,UniyalDihingiaMizunoRezzolla2026}.

Against this background, we develop a unified analytic description of the
polar critical curve and photon-ring observables in a radial separable sector
beyond Kerr, while retaining a form that permits direct comparison with the
Kerr geometry. We adopt the radial functional structure $A_1(r)$, $A_2(r)$,
$A_5(r)$, and $f(r)$ of the Johannsen separable metric \cite{Johannsen2015},
with the angular deformation functions fixed to their Kerr values. These
quantities are retained as general radial functions rather than being
restricted to expansions compatible with the parametrized post-Newtonian
(PPN) framework commonly used in the Johannsen parametrization;
PPN-motivated or finite-order forms remain admissible as particular choices.
Within this radial separable class, we organize the metric dependence in two
complementary ways. The combinations $\calP$, $\calQ$, and $\calW$ are
tailored to the separated null-geodesic equations, whereas the Kerr-relative
metric-deformation functions $\{h,f,a_1,a_2,a_5\}$ make departures from the
Kerr reference geometry explicit and provide a direct interface for concrete
Boyer--Lindquist-like metrics. The SBG variables instead emphasize the
geometric sectors of the same radial structure
\cite{SalehiBroderickGeorgiev2024}. Once a concrete metric is mapped into
these functions, the exact polar observables follow by direct substitution,
while the same dictionary supplies the variables used in the weak-deviation
expansion about Kerr. Appendix~\ref{app:metric-dictionary} collects
representative mappings. Thus the exact formulation is retained first, and
for weak departures from Kerr the same representation provides a systematic
first-order expansion about a black hole with the same asymptotic mass and
spin. The Kerr spin itself is not expanded. Because the angular sector is
held fixed to Kerr, more general separable geometries with genuine angular
deformations, such as NUT-charged or magnetized rotating spacetimes, lie
outside the present class
\cite{GrenzebachPerlickLammerzahl2014,ZhangJiang2021,
	WuWei2023,WanZhangWeiHouChen2026}.

We focus on polar observers, for whom axial symmetry reduces the critical
curve to a circle, while the asymptotic high-order image sequence is
characterized by three scalar critical parameters
$(\tilde\gamma,\tilde\tau,\tilde\delta)$. These parameters describe the demagnification, time delay, and
azimuthal rotation between successive high-order images
\cite{GrallaLupsasca2020}. Their counterparts beyond Kerr have also
been studied in concrete separable geometries, including Kerr--Newman
and Kerr--Bertotti--Robinson backgrounds
\cite{WaliaKocherlakotaChangSalehi2025,
	HouLiuGuoYanChen2022,WanZhangWeiHouChen2026}.
Within the present framework, we express the exact polar critical-curve radius
and the three critical parameters in a common representation, with
$(\tilde\gamma,\tilde\tau,\tilde\delta)$ normalized per polar half-libration,
and then derive their first-order variations for weak deviations from Kerr.
The resulting formulas identify which local radial deformation data are
sampled by each polar observable. Representative metric families are used to
test the linear approximation and illustrate this metric dependence.

The paper is organized as follows. Section~\ref{sec:metric} introduces the optical functions, their Kerr-relative metric representation, and the separated null geodesics in differential and integral forms. Section~\ref{sec:exact-polar-observables} collects the exact polar critical curve and photon-ring critical parameters and relates them to the asymptotic high-order image sequence. Section~\ref{sec:weak-deviations-from-kerr} develops their weak-deviation expansion about Kerr. Section~\ref{sec:examples} illustrates representative metric realizations, individual first-order variations, and their joint behavior in observable space. Section~\ref{sec:discussion} summarizes the main results and their relation to other separable and perturbative frameworks. Appendices~\ref{app:metric-dictionary} and \ref{app:useful-formulas} provide metric mappings and formulas useful for direct application.

Throughout this paper we use geometrized units. Tildes denote
quantities associated with a critical photon orbit, a subscript $K$
denotes the corresponding Kerr reference quantity, a bracketed
superscript $[1]$ denotes the first-order metric variation, and a prime
denotes differentiation with respect to the radial coordinate.

\section{Metric representation and null geodesics}
\label{sec:metric}

\subsection{Optical functions and Kerr-relative metric representation}
\label{subsec:metric-functions-and-basis}

We work in Boyer--Lindquist-like coordinates $(t,r,\theta,\phi)$, with mass $M$ and spin parameter $a=J/M$. As an explicit realization of this class, we use the Johannsen separable
metric restricted to the radial functions $A_1(r)$, $A_2(r)$, $A_5(r)$, and $f(r)$, with the angular deformation functions fixed to their Kerr values \cite{Johannsen2015}. The covariant line element is
\begin{align}
ds^2={}&
-\frac{\Sigma\left[\Delta_{\K}-a^2A_2^2(r)\sin^2\theta\right]}
{\left[(r^2+a^2)A_1(r)-a^2A_2(r)\sin^2\theta\right]^2}\,dt^2
\nonumber\\
&-\frac{2a\Sigma\sin^2\theta
\left[(r^2+a^2)A_1(r)A_2(r)-\Delta_{\K}\right]}
{\left[(r^2+a^2)A_1(r)-a^2A_2(r)\sin^2\theta\right]^2}\,dt\,d\phi
\nonumber\\
&+\frac{\Sigma\sin^2\theta
\left[(r^2+a^2)^2A_1^2(r)-a^2\Delta_{\K}\sin^2\theta\right]}
{\left[(r^2+a^2)A_1(r)-a^2A_2(r)\sin^2\theta\right]^2}\,d\phi^2
\nonumber\\
&+\frac{\Sigma}{\Delta_{\K}A_5(r)}\,dr^2
+\Sigma\,d\theta^2 .
\label{eq:metric-line-element}
\end{align}
Here
\begin{equation}
\Sigma(r,\theta)=r^2+a^2\cos^2\theta+f(r),
\quad
\Delta_{\K}(r)=r^2-2Mr+a^2.
\label{eq:metric-radial-functions}
\end{equation}
Setting $A_1=A_2=A_5=1$ and $f=0$ reduces Eq.~\eqref{eq:metric-line-element} to the Kerr metric in Boyer--Lindquist coordinates.
The original Johannsen parametrization usually expands the radial
functions in a PPN-compatible series. Here $A_1$, $A_2$, $A_5$, and
$f$ are instead treated as general radial functions, without assuming
a particular finite-order truncation. Throughout, we restrict attention to asymptotically flat realizations
for which $M$ and $J=aM$ are the asymptotic mass and angular momentum,
so that the standard observer-screen normalization at infinity applies.
 
The combinations entering the inverse metric and the separated null
equations are made explicit by defining two optical functions
\begin{equation}
\begin{aligned}
\calP(r)&\equiv
\frac{(r^2+a^2)A_1(r)}{\sqrt{\Delta_{\K}(r)}},
&\qquad
\calQ(r)&\equiv
\frac{aA_2(r)}{\sqrt{\Delta_{\K}(r)}},&&
\end{aligned}
\label{eq:metric-optical-functions}
\end{equation}
and a radial kinetic factor
\begin{equation}
\calW(r)\equiv \Delta_{\K}(r)A_5(r).
\label{eq:metric-kinetic-function}
\end{equation}
The inverse metric then takes the compact separable form
\begin{align}
\Sigma g^{tt}&=-\calP^2+a^2\sin^2\theta,
\nonumber\\
\Sigma g^{t\phi}&=a-\calP\calQ,
\nonumber\\
\Sigma g^{\phi\phi}&=\csc^2\theta-\calQ^2,
\nonumber\\
\Sigma g^{rr}&=\calW,
\qquad
\Sigma g^{\theta\theta}=1.
\label{eq:metric-inverse-optical}
\end{align}
Thus the null dynamics singles out $\calP$, $\calQ$, and $\calW$. The first two determine the optical combinations entering the critical impact parameters, while $\calW$ fixes the radial kinetic normalization. 
%The physical radial instability is governed by the complete combination of $\calW$ with the radial optical curvature.

The nonperturbative and nonparametric formulation of Salehi, Broderick, and Georgiev (SBG) describes the same inverse-metric structure using four radial functions $N(r)$, $F(r)$, $B(r)$, and $f_{\rm SBG}(r)$ \cite{SalehiBroderickGeorgiev2024}. Direct comparison gives
\begin{equation}
\begin{aligned}
\calP(r)&=\frac{r}{N(r)},
&\qquad
\calQ(r)&=\frac{aF(r)}{N(r)},
\\
\calW(r)&=\frac{r^2N^2(r)}{B^2(r)},
&
 f(r)&=f_{\rm SBG}(r).
\end{aligned}
\label{eq:metric-SBG-general-map}
\end{equation}
The functions $\calP$, $\calQ$, and $\calW$ are therefore precisely
the combinations of the SBG radial functions selected by the
separated photon dynamics. The SBG representation provides a complementary geometric organization: $N$ directly characterizes the horizon structure, $F$ controls the stationary--azimuthal sector and the ergoregion once $N$ is specified, $B$ fixes the radial normalization, and $f_{\rm SBG}$ enters the common conformal factor \cite{SalehiBroderickGeorgiev2024}. The optical representation instead packages these functions into the combinations that enter the separated geodesic equations, providing a convenient basis for the calculations below.\footnote{As noted in Ref.~\cite{SalehiWaliaChangKocherlakota2025}, the SBG function $B$ can be removed from the radial geodesic equation by a redefinition of the radial coordinate. Accordingly, $B$, $\calW$, or $A_5$ should not by themselves be assigned coordinate-invariant physical
significance. We keep the Boyer--Lindquist-like radial coordinate fixed throughout, for which matching to concrete metrics is direct; the physical Lyapunov response is carried by the complete radial-instability combination entering $\tilde\gamma$.}

In the Kerr limit, the two optical functions and the radial kinetic factor reduce to
\begin{equation}
\begin{aligned}
\calP_{\K}(r)&\equiv
\frac{r^2+a^2}{\sqrt{\Delta_{\K}(r)}},
&\qquad
\calQ_{\K}(r)&\equiv
\frac{a}{\sqrt{\Delta_{\K}(r)}},
\\
\calW_{\K}(r)&\equiv\Delta_{\K}(r).&&
\end{aligned}
\label{eq:metric-kerr-optical-functions}
\end{equation}
For the general metric, these definitions give the exact Kerr-factorized form
\begin{equation}
\calP=\calP_{\K}A_1,
\qquad
\calQ=\calQ_{\K}A_2,
\qquad
\calW=\calW_{\K}A_5.
\label{eq:metric-exact-optical-factorization}
\end{equation}
The functions $A_i$ are therefore exact multipliers of the corresponding Kerr optical and radial kinetic factors.

For mapping a concrete metric and for studying departures from Kerr, it is useful to isolate a generalized radial function $\Delta(r)$. Many Kerr-like metrics are naturally presented with a modified Kerr $\Delta$-function, so we introduce
\begin{equation}
\Delta(r)=\Delta_{\K}(r)+h(r),
\label{eq:metric-delta-split}
\end{equation}
and define the Kerr-relative metric-deformation functions $a_i(r)$ by
\begin{equation}
\begin{aligned}
A_1(r)&=a_1(r)\sqrt{\frac{\Delta_{\K}(r)}{\Delta(r)}},
&\qquad
A_2(r)&=a_2(r)\sqrt{\frac{\Delta_{\K}(r)}{\Delta(r)}},
\\
A_5(r)&=a_5(r)\frac{\Delta(r)}{\Delta_{\K}(r)}.
\end{aligned}
\label{eq:metric-johannsen-map}
\end{equation}
Consequently,
\begin{equation}
\begin{aligned}
\calP(r)&=\frac{(r^2+a^2)a_1(r)}{\sqrt{\Delta(r)}},
&\qquad
\calQ(r)&=\frac{a\,a_2(r)}{\sqrt{\Delta(r)}},
\\
\calW(r)&=\Delta(r)a_5(r).&&
\end{aligned}
\label{eq:metric-kerr-adapted-optical-functions}
\end{equation}
A given metric does not uniquely fix $\Delta$ and $a_i$ separately;
different choices of $\Delta$ merely redistribute the same metric
information between $\Delta$ and the $a_i$. The convention adopted for each mapping
is specified explicitly in Appendix~\ref{app:metric-dictionary}.
With this convention fixed, the five functions
$\{h,f,a_1,a_2,a_5\}$ provide a Kerr-relative metric-deformation
dictionary for concrete metrics.
Equation~\eqref{eq:metric-kerr-adapted-optical-functions} then provides
$\calP$, $\calQ$, and $\calW$ directly, and substitution back into the
compact inverse metric in Eq.~\eqref{eq:metric-inverse-optical} makes
the corresponding deviations from Kerr manifest. This representation
is therefore ready for direct use in the exact formulas of
Sec.~\ref{sec:exact-polar-observables}, while its weak-deviation limit
leads immediately to the first-order deformation profiles $U_i$ introduced in
Sec.~\ref{sec:weak-deviations-from-kerr}. The Kerr limit is obtained
by setting $f=h=0$ and all $a_i$ to unity.

\subsection{Separated null-geodesic equations}
\label{subsec:null-geodesics}

The metric introduced above admits a separable Hamilton--Jacobi equation for null geodesics \cite{Johannsen2015,SalehiBroderickGeorgiev2024}. We use
\begin{equation}
S=-Et+L\phi+S_r(r)+S_\theta(\theta),
\label{eq:null-geodesics-hj-ansatz}
\end{equation}
where $E=-p_t$ and $L=p_\phi$ are the conserved energy and axial angular momentum. We introduce the energy-rescaled constants
\begin{equation}
\xi=\frac{L}{E},
\qquad
\eta=\frac{\mathcal K}{E^2},
\label{eq:null-geodesics-impact-parameters}
\end{equation}
and
\begin{equation}
\zeta^2=\eta+(\xi-a)^2,
\label{eq:null-geodesics-zeta}
\end{equation}
where $\mathcal K$ is the separation constant. The radial optical combination appearing in the separated equations is
\begin{equation}
\calH(r;\xi)=\calP(r)-\xi\calQ(r).
\label{eq:null-geodesics-H}
\end{equation}

Let $\lambda$ be an affine parameter and introduce the Mino-type parameter $\nu$ by
\begin{equation}
\dd\nu=\frac{E}{\Sigma}\,\dd\lambda.
\label{eq:high-order-rings-mino-parameter}
\end{equation}
The separated first-order null equations then take the compact form
\begin{align}
\frac{dt}{d\nu}
&=\calP\calH+a\left(\xi-a\sin^2\theta\right),
\label{eq:null-geodesics-t-equation}
\\
\frac{dr}{d\nu}
&=\pm_r\sqrt{\calR(r)},
\label{eq:null-geodesics-r-equation}
\\
\frac{d\theta}{d\nu}
&=\pm_\theta\sqrt{\Theta(\theta)},
\label{eq:null-geodesics-theta-equation}
\\
\frac{d\phi}{d\nu}
&=\calQ\calH+\xi\csc^2\theta-a.
\label{eq:null-geodesics-phi-equation}
\end{align}
Here $\pm_r$ and $\pm_\theta$ denote the signs of the radial and polar velocities. 
The radial and angular potentials are
\begin{equation}
	\calR(r)=\calW(r)\left[\calH^2(r;\xi)-\zeta^2\right],
	\label{eq:null-geodesics-radial-potential}
\end{equation}
and
\begin{equation}
	\Theta(\theta)=\eta+a^2\cos^2\theta-\xi^2\cot^2\theta.
	\label{eq:null-geodesics-angular-potential}
\end{equation}
The designation ``Mino-type'' emphasizes that $\nu$ is adapted to the
separable radial sector considered here and should not be confused with
the affine parameter $\lambda$.
These equations make the metric dependence explicit. The position of a critical orbit is controlled by the optical functions, while $\calW$ multiplies the radial potential as a kinetic factor. The function $f(r)$ has been removed from the separated equations by the Mino-type reparametrization and therefore does not change the critical impact parameters. Likewise, a finite and nonzero $\calW$ does not move the double root, although it contributes to the complete radial-instability combination entering the high-order photon-ring exponent.

\subsection{Null geodesics in integral form}
\label{subsec:null-geodesics-integral-form}

For later use, it is convenient to separate the purely radial pieces of the temporal and azimuthal equations,
\begin{equation}
\begin{aligned}
\mathcal T_r(r;\xi)
&\equiv
\calP(r)\calH(r;\xi)+a\xi-a^2,
\\
\Phi_r(r;\xi)
&\equiv
\calQ(r)\calH(r;\xi)-a.
\end{aligned}
\label{eq:null-geodesics-radial-time-azimuth-functions}
\end{equation}
Equations~\eqref{eq:null-geodesics-t-equation} and \eqref{eq:null-geodesics-phi-equation} then read
\begin{equation}
\frac{dt}{d\nu}=\mathcal T_r+a^2\cos^2\theta,
\qquad
\frac{d\phi}{d\nu}=\Phi_r+\xi\csc^2\theta.
\label{eq:null-geodesics-separated-time-azimuth}
\end{equation}
Following the standard Kerr-lensing convention \cite{GrallaLupsasca2020,SalehiWaliaChangKocherlakota2025}, define the path-dependent radial integrals
\begin{equation}
\begin{aligned}
I_t&\equiv\int_{\rm path}
\frac{\mathcal T_r(r;\xi)}{\pm_r\sqrt{\calR(r)}}\,\dd r,
&\qquad
I_r&\equiv\int_{\rm path}
\frac{\dd r}{\pm_r\sqrt{\calR(r)}},
\\
I_\phi&\equiv\int_{\rm path}
\frac{\Phi_r(r;\xi)}{\pm_r\sqrt{\calR(r)}}\,\dd r,
\end{aligned}
\label{eq:null-geodesics-radial-integrals}
\end{equation}
and the angular integrals
\begin{equation}
\begin{aligned}
G_t&\equiv\int_{\rm path}
\frac{\cos^2\theta}{\pm_\theta\sqrt{\Theta(\theta)}}\,\dd\theta,
&\qquad
G_\theta&\equiv\int_{\rm path}
\frac{\dd\theta}{\pm_\theta\sqrt{\Theta(\theta)}},
\\
G_\phi&\equiv\int_{\rm path}
\frac{\csc^2\theta}{\pm_\theta\sqrt{\Theta(\theta)}}\,\dd\theta.
\end{aligned}
\label{eq:null-geodesics-angular-integrals}
\end{equation}
The signs change at the corresponding turning points, so each path integral is understood as the sum over monotonic radial or polar segments. Integrating the separated equations along a null trajectory gives the endpoint increments
\begin{equation}
\Delta t=I_t+a^2G_t,
\qquad
\Delta\phi=I_\phi+\xi G_\phi,
\qquad
\Delta\nu=I_r=G_\theta.
\label{eq:null-geodesics-integral-form}
\end{equation}
This representation makes the radial--angular separation used below explicit. Because the angular potential retains its Kerr form, the angular integrals have the standard Kerr elliptic structure, whereas the radial integrals carry the spacetime dependence through $\calP$, $\calQ$, and $\calW$.

\section{Polar critical curve and photon-ring critical parameters}
\label{sec:exact-polar-observables}
\label{sec:polar-critical-curve}

\subsection{Observer screen}
\label{subsec:polar-observer-screen}

We first recall the map from impact parameters to the observer screen.  For an asymptotic observer at a generic inclination $\theta_o$, the standard celestial coordinates are \cite{Bardeen1973,PerlickTsupko2022}
\begin{equation}
\alpha=-\frac{\xi}{\sin\theta_o},
\label{eq:polar-screen-alpha-general}
\end{equation}
and
\begin{equation}
\beta=\pm\sqrt{\eta+a^2\cos^2\theta_o-\xi^2\cot^2\theta_o}.
\label{eq:polar-screen-beta-general}
\end{equation}
These coordinates are useful away from the axis.  The polar limit, however, should not be obtained by a naive substitution $\theta_o=0$, because the Cartesian screen direction becomes degenerate on the rotation axis.

For a polar observer, axial symmetry reduces the critical curve to a circle. We therefore use image-plane polar coordinates $(\rho,\varphi)$, where $\rho$ is the screen radius and $\varphi$ is the screen azimuthal angle. Regular null rays reaching the rotation axis have zero axial angular momentum, and the screen relation becomes
\begin{equation}
\xi=0,
\qquad
\rho^2=\eta+a^2=\zeta^2,
\qquad
\rho=\zeta>0.
\label{eq:polar-screen-xi-zero}
\end{equation}

\subsection{Critical impact parameters}
\label{subsec:spherical-critical-impact-parameters}

Spherical photon orbits and their relation to separability in Kerr and non-Kerr spacetimes have been studied in Refs.~\cite{Teo2003,GlampedakisPappas2019,MengFanLiHanZhang2023}. Broader existence and stability properties of light rings in stationary and axisymmetric spacetimes were established in Ref.~\cite{GuoGao2021}. A spherical photon orbit at $r=\tilde r$ is determined by
\begin{equation}
\calR(\tilde r)=0,
\qquad
\calR'(\tilde r)=0.
\label{eq:null-geodesics-critical-conditions-R}
\end{equation}
Since $\calW$ is assumed to be finite and nonzero on the orbit, these conditions are equivalently
\begin{equation}
\calH^2(\tilde r;\tilde\xi)=\tilde\zeta^2,
\qquad
\left[\calH^2(r;\tilde\xi)-\tilde\zeta^2\right]'_{r=\tilde r}=0.
\label{eq:null-geodesics-critical-conditions-H}
\end{equation}
On the ordinary spherical photon-orbit branch, $\calH(\tilde r;\tilde\xi)\ne0$, and hence
\begin{equation}
\tilde\xi
=
\frac{\calP'(\tilde r)}{\calQ'(\tilde r)},
\qquad
\tilde\zeta
=
\calP(\tilde r)-\tilde\xi\calQ(\tilde r),
\label{eq:null-geodesics-critical-xi-zeta-rescaled}
\end{equation}
for $\calQ'(\tilde r)\ne0$, with the sign of $\tilde\zeta$ chosen positive on the exterior branch. The remaining impact parameter is
\begin{equation}
\tilde\eta
=
\tilde\zeta^2-(\tilde\xi-a)^2.
\label{eq:null-geodesics-critical-eta-rescaled}
\end{equation}
These relations describe the spherical critical family before specifying the observer inclination. They depend on the two optical functions $\calP$ and $\calQ$, whereas the radial kinetic factor $\calW$ does not shift the double root.

\subsection{Polar critical curve}
\label{subsec:polar-critical-curve}

For a polar observer, regular rays reaching the rotation axis satisfy
$\tilde\xi=0$. Substituting this condition directly into the double-root
conditions in Eq.~\eqref{eq:null-geodesics-critical-conditions-H} gives,
on the ordinary exterior branch,
\begin{equation}
	\calP'(\tilde r)=0.
	\label{eq:polar-exact-branch-critical-data}
\end{equation}
Combining the remaining spherical-orbit relation with the polar screen map in Eq.~\eqref{eq:polar-screen-xi-zero}, the critical curve is a circle of radius
\begin{equation}
\tilde b\equiv\tilde\rho=\tilde\zeta
=\calP(\tilde r).
\label{eq:polar-exact-branch-R-exact}
\end{equation}
The polar critical orbit is therefore the stationary point of $\calP$, and the screen radius is its stationary value.

Using Eqs.~\eqref{eq:metric-exact-optical-factorization}--\eqref{eq:metric-kerr-adapted-optical-functions}, the exact radius can be written directly as
\begin{equation}
\tilde b
=
\calP_{\K}(\tilde r)\,A_1(\tilde r)
=
\calP_{\K}(\tilde r)\,a_1(\tilde r)
\left[1+\frac{h(\tilde r)}{\Delta_{\K}(\tilde r)}\right]^{-1/2}.
\label{eq:polar-exact-branch-R-h}
\end{equation}
In general the deformed critical-orbit coordinate $\tilde r$ differs from its Kerr value $\tilde r_{\K}$. Accordingly, $\calP_{\K}(\tilde r)=(\tilde r^2+a^2)/\sqrt{\Delta_{\K}(\tilde r)}$ is the Kerr optical factor evaluated at the deformed orbit coordinate and should not be identified with the Kerr critical-curve radius $\tilde b_{\K}=\calP_{\K}(\tilde r_{\K})$.

Using the Kerr-relative metric-deformation representation, the stationary condition in Eq.~\eqref{eq:polar-exact-branch-critical-data} can equivalently be written as
\begin{equation}
0=
\left.\frac{\calP'}{\calP}\right|_{r=\tilde r}
=
\left[
\frac{2r}{r^2+a^2}
+
\frac{a_1'}{a_1}
-
\frac{1}{2}
\frac{\Delta_{\K}'+h'}{\Delta_{\K}+h}
\right]_{r=\tilde r}.
\label{eq:polar-exact-branch-log-condition-h}
\end{equation}
This form allows the polar critical orbit to be obtained directly once $\Delta(r)$ and $a_1(r)$ are identified for a concrete metric.

\paragraph*{Metric dependence of the polar critical curve.}
At the level of the optical functions, the polar critical orbit and its screen radius depend only on $\calP$: the condition $\tilde\xi=0$ removes $\calQ$ from $\calH=\calP-\xi\calQ$, while a finite and nonzero $\calW$ multiplies the radial potential without shifting its double root. The function $f$ is absent from the separated null equations after the Mino-type reparametrization. In the Kerr-relative metric-deformation representation this means that the polar critical curve is controlled by the combined $(\Delta,a_1)$ dependence of $\calP$, whereas $a_2$, $a_5$, and $f$ do not change $\tilde b$. The additional photon-ring parameters probe complementary information: the Lyapunov exponent is sensitive to the complete radial-instability combination, and the rotation parameter directly samples the second optical function $\calQ$.

\subsection{Polar photon-ring parameters}
\label{sec:high-order-polar-photon-rings}
\label{subsec:high-order-critical-rays-exact-lyapunov}

On the polar critical orbit determined in Sec.~\ref{subsec:polar-critical-curve}, the conserved quantities take $\tilde\xi=0$ and $\tilde\zeta=\tilde b$. The corresponding radial and angular potentials reduce to
\begin{equation}
\calR_{\rm c}(r)
=\calW(r)\left[\calP^2(r)-\tilde b^2\right],
\qquad
\Theta_{\rm c}(\theta)
=\tilde b^2-a^2\sin^2\theta.
\label{eq:high-order-rings-critical-potentials}
\end{equation}
Here $\calP(\tilde r)=\tilde b$ and $\calP'(\tilde r)=0$ ensure the radial double root. All three critical parameters below are normalized per polar half-libration, from one rotation axis to the other, following the standard photon-ring convention \cite{GrallaLupsasca2020,SalehiWaliaChangKocherlakota2025}. 
%Here $K(m)$ and $E(m)$ denote the complete Legendre elliptic integrals of the first and second kinds, respectively. Their argument $m$ is the elliptic parameter; the corresponding elliptic modulus is $k=\sqrt{m}$.

\paragraph*{Lyapunov exponent.}
For a small radial displacement $r(\nu)=\tilde r+\delta r(\nu)$, the double-root expansion gives
\begin{equation}
\frac{\dd\,\delta r}{\dd\nu}
=\pm\tilde\kappa\,\delta r+O(\delta r^2),
\qquad
\tilde\kappa^2
=\frac{1}{2}\calR_{\rm c}''(\tilde r).
\label{eq:high-order-rings-radial-instability-rate}
\end{equation}
Because both $\calP^2-\tilde b^2$ and its first derivative vanish at the critical orbit, derivatives of $\calW$ do not contribute, and
\begin{equation}
\tilde\kappa
=
\left[
\calW(\tilde r)\tilde b\,\calP''(\tilde r)
\right]^{1/2}.
\label{eq:high-order-rings-kappa-explicit}
\end{equation}
The angular integral measuring the Mino-type duration of one polar half-libration is
\begin{equation}
\tilde G_\theta
\equiv
\int_0^\pi\frac{\dd\theta}{\sqrt{\Theta_{\rm c}(\theta)}}
=
\frac{2}{\tilde b}K(\mu),
\qquad
\mu\equiv\frac{a^2}{\tilde b^2}.
\label{eq:high-order-rings-polar-half-libration}
\end{equation}
Here and below, $K(m)$ and $E(m)$ denote the complete elliptic integrals
of the first and second kinds in the parameter convention $m=k^2$.
The dimensionless Lyapunov exponent is the radial instability accumulated over this angular half-libration,
\begin{equation}
\tilde\gamma
\equiv
\tilde\kappa\tilde G_\theta
=
\frac{2}{\tilde b}K(\mu)
\left[
\calW(\tilde r)\tilde b\,\calP''(\tilde r)
\right]^{1/2}.
\label{eq:high-order-rings-lyapunov-explicit}
\end{equation}
Because the intermediate instability rate $\tilde\kappa$ depends on the
choice of evolution parameter, it must be combined with the polar
half-libration interval defined using the same parametrization. The
resulting quantity $\tilde\gamma=\tilde\kappa\tilde G_\theta$ is the
dimensionless instability exponent entering the asymptotic photon-ring
scaling \cite{DeichYunesGammie2024}. Related analyses connect photon-ring or light-ring instability with wave dynamics and quasinormal modes \cite{GiataganasKehagiasRiotto2024,GuoZhongWangGao2022,Li:2021zct}. The radial instability entering $\tilde\gamma$ is governed by the
complete combination $\calW \calP''$ rather than by
$\calW$ alone; an explicit expression for $\calP''$ in terms of $\Delta$ and $a_1$ is collected in Appendix~\ref{app:useful-formulas}.

\paragraph*{Time delay.}
The second angular integral entering the temporal equation is
\begin{equation}
\tilde G_t
\equiv
\int_0^\pi
\frac{\cos^2\theta\,\dd\theta}{\sqrt{\Theta_{\rm c}(\theta)}}
=
\frac{2}{\mu\,\tilde b}
\left[E(\mu)-(1-\mu)K(\mu)\right].
\label{eq:high-order-rings-polar-time-integral}
\end{equation}
From Eq.~\eqref{eq:null-geodesics-integral-form}, the coordinate-time increment over the same half-libration is
\begin{align}
\tilde\tau
&=
\mathcal T_r(\tilde r;0)\tilde G_\theta+a^2\tilde G_t
\nonumber\\
&=(\tilde b^2-a^2)\tilde G_\theta+a^2\tilde G_t
=2\tilde b\,E(\mu).
\label{eq:high-order-rings-time-delay-exact}
\end{align}
Here $\tilde\tau$ is a Boyer--Lindquist coordinate-time interval, not a
Mino-time period. With the asymptotically flat normalization adopted
here, this coordinate time agrees with the time measured by a static
observer at infinity.
Thus, within the present radial sector, $\tilde\tau$ is fixed by the polar critical-curve radius and the spin and introduces no additional independent radial metric function.

\paragraph*{Rotation parameter.}
The third angular integral is
\begin{equation}
G_\phi(\xi)
\equiv
\int_{\theta_-}^{\theta_+}
\frac{\csc^2\theta\,\dd\theta}{\sqrt{\Theta(\theta;\xi)}}.
\label{eq:high-order-rings-near-polar-Gphi}
\end{equation}
For nonzero $\xi$, it can be represented by the complete Legendre elliptic integral of the third kind $\Pi(n|m)$. Its general closed form is not needed here. The azimuthal advance of a nearby spherical orbit over one polar half-libration is
\begin{equation}
\Delta\phi_{1/2}(\xi)
=
\Phi_r(\tilde r;\xi)G_\theta(\xi)+\xi G_\phi(\xi).
\label{eq:high-order-rings-near-polar-azimuthal-advance}
\end{equation}
Although $G_\phi$ itself diverges as the turning points approach the axes, the combination entering the azimuthal advance has the finite polar limit \cite{GrallaLupsasca2020}
\begin{equation}
\lim_{\xi\to0}\xi G_\phi(\xi)=\pi.
\label{eq:high-order-rings-xi-Gphi-limit}
\end{equation}
The continuous rotation parameter is therefore
\begin{align}
\tilde\delta
&=
\Phi_r(\tilde r;0)\tilde G_\theta
+\lim_{\xi\to0}\xi G_\phi(\xi)
\nonumber\\
&=
\pi+
\left[\tilde b\,\calQ(\tilde r)-a\right]\tilde G_\theta
\nonumber\\
&=
\pi+2K(\mu)
\left[\calQ(\tilde r)-\frac{a}{\tilde b}\right].
\label{eq:high-order-rings-rotation-exact}
\end{align}
The corresponding screen rotation may be reduced modulo $2\pi$,
\begin{equation}
\tilde\delta_{2\pi}=\tilde\delta\bmod 2\pi,
\label{eq:high-order-rings-rotation-wrapped}
\end{equation}
but the continuous quantity is the appropriate variable for analytic continuation and perturbative differentiation. Unlike $\tilde b$, $\tilde\gamma$, and $\tilde\tau$, the rotation parameter directly samples the second optical function $\calQ=\calQ_{\K}A_2$, or equivalently $\calQ=aa_2/\sqrt{\Delta}$ in the Kerr-relative metric representation.

\paragraph*{Exact polar observable hierarchy.}
The four polar quantities resolve three distinct combinations of the radial geometry. The pair $(\tilde b,\tilde\tau)$ is fixed by the stationary optical value $\calP(\tilde r)$ and the spin; $\tilde\gamma$ additionally samples the complete radial-instability combination $\calW(\tilde r)\calP''(\tilde r)$; and $\tilde\delta$ additionally samples $\calQ(\tilde r)$. The function $f(r)$ remains an exact blind direction of this polar null-geodesic data set in the radial sector considered here.

\subsection{Connection to the asymptotic high-order image sequence}
\label{subsec:high-order-near-critical-rays-photon-rings}

The quantities $(\tilde\gamma,\tilde\tau,\tilde\delta)$ are exact properties of the polar critical orbit. Their asymptotic interpretation in successive high-order images follows the standard critical-parameter construction \cite{GrallaLupsasca2020,SalehiWaliaChangKocherlakota2025}: nearly critical rays spend an increasing number of polar half-librations near $r=\tilde r$ before reaching the observer. Let
\begin{equation}
d_m\equiv\rho_m-\tilde b
\label{eq:high-order-rings-screen-displacement}
\end{equation}
be the signed displacement of the $m$th member of a high-order image sequence from the critical circle, where $m$ increases by one per polar half-libration. For each asymptotic image family,
\begin{equation}
d_m=C_\pm e^{-m\tilde\gamma}\left[1+o(1)\right],
\qquad m\to\infty,
\label{eq:high-order-rings-asymptotic-displacement}
\end{equation}
with a source- and family-dependent coefficient $C_\pm$. Consequently,
\begin{equation}
\lim_{m\to\infty}
\left|\frac{d_{m+1}}{d_m}\right|
=e^{-\tilde\gamma},
\label{eq:high-order-rings-scaling-law}
\end{equation}
while the successive arrival-time and orientation increments obey
\begin{align}
\lim_{m\to\infty}(t_{m+1}-t_m)
&=\tilde\tau,
\nonumber\\
\lim_{m\to\infty}(\varphi_{m+1}-\varphi_m)
&=\tilde\delta\pmod{2\pi}.
\label{eq:high-order-rings-tau-delta-scaling}
\end{align}
Thus the source- and family-dependent prefactors and offsets cancel in
these leading relations between successive image orders. If the image order is instead incremented only after a complete polar
libration, the corresponding parameters are $(2\tilde\gamma,
2\tilde\tau,2\tilde\delta)$, with the rotation understood modulo $2\pi$ on the screen. The half-libration convention adopted here keeps all three critical parameters normalized consistently.

\section{Weak deviations from Kerr}
\label{sec:weak-deviations-from-kerr}

The exact results of Sec.~\ref{sec:exact-polar-observables} are nonperturbative within the radial separable sector. We now expand them for weak deviations from a Kerr
reference with the same asymptotic mass and spin. Physically, this
weak-deviation expansion is intended for a near-Kerr regime in which the Kerr
geometry provides a good zeroth-order description even in the
strong-field neighborhood of the photon shell. The aim is to isolate
the local deformation data entering the critical orbit, the
critical-curve radius, and the three half-libration critical parameters.

\subsection{Kerr reference and weak-deviation variables}
\label{subsec:polar-kerr-weak-response}

The Kerr functions $\Delta_{\K}$, $\calP_{\K}$, and $\calQ_{\K}$ were defined in Sec.~\ref{subsec:metric-functions-and-basis}. The Kerr polar critical orbit $\tilde r_{\K}$ satisfies
\begin{equation}
\calP_{\K}'(\tilde r_{\K})=0,
\label{eq:polar-kerr-critical-radius-condition}
\end{equation}
or equivalently
\begin{equation}
\tilde r_{\K}^3-3M\tilde r_{\K}^2+a^2\tilde r_{\K}+Ma^2=0.
\label{eq:polar-kerr-critical-radius-cubic}
\end{equation}
The corresponding critical-curve radius is
\begin{equation}
\tilde b_{\K}
=\calP_{\K}(\tilde r_{\K})
=\frac{\tilde r_{\K}^2+a^2}{\sqrt{\Delta_{\K}(\tilde r_{\K})}}.
\label{eq:polar-kerr-critical-curve-radius}
\end{equation}
The Kerr radial instability and Lyapunov exponent are
\begin{equation}
\tilde\kappa_{\K}
=
\left[
\Delta_{\K}(\tilde r_{\K})
\tilde b_{\K}\,\calP_{\K}''(\tilde r_{\K})
\right]^{1/2},
\qquad
\tilde\gamma_{\K}=\tilde\kappa_{\K}\tilde G_{\theta\K},
\label{eq:high-order-rings-Kerr-kappa-and-gamma}
\end{equation}
where
\begin{equation}
\tilde G_{\theta\K}=\frac{2}{\tilde b_{\K}}K(\mu_{\K}),
\qquad
\mu_{\K}=\frac{a^2}{\tilde b_{\K}^2}.
\label{eq:high-order-rings-Kerr-angular-quantities}
\end{equation}
The remaining Kerr critical parameters are
\begin{align}
\tilde\tau_{\K}
&=2\tilde b_{\K}E(\mu_{\K}),
\label{eq:high-order-rings-Kerr-time-delay}\\
\tilde\delta_{\K}
&=\pi+2K(\mu_{\K})
\left[
\calQ_{\K}(\tilde r_{\K})-\frac{a}{\tilde b_{\K}}
\right].
\label{eq:high-order-rings-Kerr-rotation}
\end{align}

The weak expansion is local to the strong-field neighborhood of $\tilde r_{\K}$ and is not an expansion in $M/r$. In the Kerr-relative metric-deformation representation, write
\begin{equation}
a_i(r)=1+\epsilon_i(r),
\qquad i=1,2,5,
\label{eq:polar-kerr-weak-deviation-ai}
\end{equation}
and
\begin{equation}
\Delta(r)=\Delta_{\K}(r)\left[1+\epsD(r)\right],
\qquad
\epsD(r)=\frac{h(r)}{\Delta_{\K}(r)}.
\label{eq:polar-kerr-weak-deviation-Delta}
\end{equation}
The perturbations and the radial derivatives required below are assumed small in the relevant strong-field interval. Equivalently, the exact multipliers are written as
\begin{equation}
A_i(r)=1+U_i(r)+O(U^2),
\qquad i=1,2,5,
\label{eq:kerr-response-Ai-Ui}
\end{equation}
with
\begin{equation}
U_1=\epsilon_1-\frac12\epsD,
\qquad
U_2=\epsilon_2-\frac12\epsD,
\qquad
U_5=\epsilon_5+\epsD.
\label{eq:kerr-response-Ui-epsilon-map}
\end{equation}
Although $\epsD$ and $\epsilon_i$ separately depend on the chosen
decomposition, the combinations $U_i$ are the first-order perturbations
of the original multipliers $A_i$ and are therefore independent of this
algebraic redistribution. The $U_i$ are defined in the fixed
Boyer--Lindquist-like radial representation adopted here. They provide
a useful bookkeeping device for metric deformations, while the complete
observable variations do not depend on assigning physical meaning to an
individual radial normalization factor.

For any critical quantity used below, we write
\begin{equation}
\tilde X=\tilde X_{\K}+\tilde X^{[1]}+O(\epsilon^2),
\qquad
\tilde X\in
\{\tilde r,\tilde b,\tilde\gamma,\tilde\tau,\tilde\delta\}.
\label{eq:kerr-response-general-expansion}
\end{equation}
The superscript $[1]$ denotes the first-order change induced by the metric deformation and is distinct from the near-critical trajectory displacement $\delta r(\nu)$ introduced in Sec.~\ref{sec:high-order-polar-photon-rings}.

\subsection{Critical orbit and critical-curve radius at first order}
\label{subsec:critical-orbit-radius-responses}

The polar orbit and critical-curve radius are controlled at first order by the optical deformation $U_1$ defined in Eq.~\eqref{eq:kerr-response-Ui-epsilon-map}:
\begin{equation}
\calP(r)=\calP_{\K}(r)\left[1+U_1(r)\right]+O(\epsilon^2).
\label{eq:polar-linear-radius-response-calP}
\end{equation}
Expanding the stationary condition $\calP'(\tilde r)=0$ about $\tilde r_{\K}$ and using $\calP_{\K}'(\tilde r_{\K})=0$ gives
\begin{equation}
\tilde r^{[1]}
=-\frac{\tilde b_{\K}}{\calP_{\K}''(\tilde r_{\K})}
U_1'(\tilde r_{\K}).
\label{eq:polar-linear-radius-response-r-first-order}
\end{equation}
Thus the orbit shift probes the radial slope of the effective optical deformation.

The critical-curve radius is the stationary value $\tilde b=\calP(\tilde r)$. Its first-order expansion separates the effect of shifting the
evaluation point from the direct deformation of the function itself,
\begin{equation}
	\tilde b^{[1]}
	=
	\calP_{\K}'(\tilde r_{\K})\tilde r^{[1]}
	+\calP_{\K}(\tilde r_{\K})U_1(\tilde r_{\K}).
	\label{eq:polar-linear-radius-response-R-first-order-intermediate}
\end{equation}
For the critical-curve radius the first contribution vanishes by the Kerr stationarity condition in Eq.~\eqref{eq:polar-kerr-critical-radius-condition}. The remaining direct optical contribution is therefore
\begin{equation}
	\frac{\tilde b^{[1]}}{\tilde b_{\K}}
	=U_1(\tilde r_{\K}).
	\label{eq:polar-linear-radius-response-R-first-order}
\end{equation}
The orbit shift and radius variation therefore probe different local data of the same function: $U_1'$ and $U_1$, respectively. At this order the polar critical-curve radius is insensitive to $U_2$, $U_5$, and $f$.

\subsection{Lyapunov exponent at first order}
\label{subsec:high-order-rings-kerr-weak-response}

The factorized form $\tilde\gamma=\tilde\kappa\tilde G_\theta$ separates the angular half-libration from the radial instability. Its first-order response is
\begin{equation}
\frac{\tilde\gamma^{[1]}}{\tilde\gamma_{\K}}
=
\frac{\tilde G_{\theta}^{[1]}}{\tilde G_{\theta\K}}
+
\frac{\tilde\kappa^{[1]}}{\tilde\kappa_{\K}}.
\label{eq:high-order-rings-gamma-response-split}
\end{equation}
The angular contribution follows entirely from the response of $\tilde b$:
\begin{equation}
\frac{\tilde G_{\theta}^{[1]}}{\tilde G_{\theta\K}}
=-\frac{E(\mu_{\K})}{[1-\mu_{\K}]K(\mu_{\K})}
U_1(\tilde r_{\K}).
\label{eq:high-order-rings-Gtheta-response}
\end{equation}
At fixed radius, the optical-curvature response at the Kerr critical orbit is
\begin{equation}
\left[\calP''\right]^{[1]}(\tilde r_{\K})
=
\calP_{\K}''(\tilde r_{\K})U_1(\tilde r_{\K})
+
\tilde b_{\K}U_1''(\tilde r_{\K}).
\label{eq:high-order-rings-calP-second-derivative-response-Kerr}
\end{equation}
Expanding the radial instability at the shifted orbit gives
\begin{align}
\frac{\tilde\kappa^{[1]}}{\tilde\kappa_{\K}}
={}&
\biggl[
U_1+\frac12 U_5
+\frac{\tilde b_{\K}}{2\calP_{\K}''}U_1''
\nonumber\\
&\hspace{18mm}
+\frac{\tilde r^{[1]}}{2}
\left(
\frac{\Delta_{\K}'}{\Delta_{\K}}
+\frac{\calP_{\K}'''}{\calP_{\K}''}
\right)
\biggr]_{r=\tilde r_{\K}}.
\label{eq:high-order-rings-kappa-response}
\end{align}
Substituting Eq.~\eqref{eq:polar-linear-radius-response-r-first-order} yields the local Lyapunov response
\begin{align}
\frac{\tilde\gamma^{[1]}}{\tilde\gamma_{\K}}
={}&
\biggl[
\frac12 U_5
+\left(1-\frac{E(\mu_{\K})}{[1-\mu_{\K}]K(\mu_{\K})}\right)U_1
\nonumber\\
&+\frac{\tilde b_{\K}}{2\calP_{\K}''}
\left\{
U_1''-
\left(
\frac{\Delta_{\K}'}{\Delta_{\K}}
+\frac{\calP_{\K}'''}{\calP_{\K}''}
\right)U_1'
\right\}
\biggr]_{r=\tilde r_{\K}}.
\label{eq:high-order-local-response-gamma}
\end{align}
The elliptic weight is the angular half-libration correction, while the remaining terms resolve the local optical curvature and radial kinetic response. The Kerr derivatives appearing in this expression are collected in Appendix~\ref{app:useful-formulas}. In the adopted radial representation, the $U_5$ contribution is invisible to the polar critical-curve radius but enters the high-order photon-ring scaling; only the complete Lyapunov response should be regarded as the physical instability observable.

\subsection{Time delay and rotation parameter at first order}
\label{subsec:time-rotation-responses}

At fixed $M$ and $a$, differentiating $\tilde\tau=2\tilde bE(a^2/\tilde b^2)$ gives
\begin{align}
\tilde\tau^{[1]}
&=2K(\mu_{\K})\tilde b^{[1]},
\label{eq:high-order-local-response-tau}\\
\frac{\tilde\tau^{[1]}}{\tilde\tau_{\K}}
&=\frac{K(\mu_{\K})}{E(\mu_{\K})}
U_1(\tilde r_{\K}).
\label{eq:high-order-local-response-relative-tau}
\end{align}
Thus the time-delay and critical-curve radius responses are proportional at
fixed spin; $\tilde\tau$ carries a different elliptic weight but no additional local radial function.

For the rotation parameter, the shifted-orbit response of the second optical function is
\[
[\calQ(\tilde r)]^{[1]}
=\calQ_{\K}(\tilde r_{\K})U_2(\tilde r_{\K})
+\calQ_{\K}'(\tilde r_{\K})\tilde r^{[1]}.
\]
Expanding Eq.~\eqref{eq:high-order-rings-rotation-exact}, using Eq.~\eqref{eq:polar-linear-radius-response-r-first-order} and the standard identity $\dd K/\dd\mu=E/[2\mu(1-\mu)]-K/(2\mu)$, gives a result involving only $K$ and $E$,
\begin{align}
\tilde\delta^{[1]}
={}&2K(\mu_{\K})\left[\calQ_{\K}U_2-\frac{\tilde b_{\K}\calQ_{\K}'}{\calP_{\K}''}U_1'+\frac{a}{\tilde b_{\K}}U_1\right]
\nonumber\\
&-2\left[\frac{E(\mu_{\K})}{1-\mu_{\K}}-K(\mu_{\K})\right]\left(\calQ_{\K}-\frac{a}{\tilde b_{\K}}\right)U_1.
\label{eq:high-order-local-response-delta}
\end{align}
Here all unmarked radial functions are evaluated at $r=\tilde r_{\K}$.
The rotation response therefore probes the second optical deformation $U_2$, together with the value and slope of $U_1$. It is independent of $U_5$ and of the conformal function $f$.

\subsection{Local metric dependence and independent polar information}
\label{subsec:local-response-hierarchy}

The preceding results reveal a local hierarchy of metric dependence on the Kerr photon shell. Here ``local'' means that the first-order observables depend only on the values and a finite number of radial derivatives of the deformation functions at the Kerr polar critical orbit $r=\tilde r_{\K}$, rather than on their complete radial profiles. With all profiles evaluated at this radius,
\begin{equation}
\begin{aligned}
\tilde r^{[1]}&\longleftrightarrow U_1',\\
\left(\frac{\tilde b^{[1]}}{\tilde b_{\K}},
\frac{\tilde\tau^{[1]}}{\tilde\tau_{\K}}\right)
&\longleftrightarrow U_1,\\
\frac{\tilde\gamma^{[1]}}{\tilde\gamma_{\K}}
&\longleftrightarrow (U_1,U_1',U_1'',U_5),\\
\tilde\delta^{[1]}
&\longleftrightarrow (U_1,U_1',U_2).
\end{aligned}
\label{eq:local-response-hierarchy}
\end{equation}
The hierarchy is then transparent. The critical-curve radius
and time delay depend on the local optical amplitude $U_1$, while the
shift of the critical orbit probes its first radial derivative $U_1'$.
The Lyapunov exponent extends this sensitivity to $U_1''$ and to the
radial kinetic deformation $U_5$ in the fixed radial representation. The
rotation parameter instead adds sensitivity to the second optical
deformation $U_2$, while also depending on $U_1$ and $U_1'$. Thus, at
first order, the polar critical data sample only a finite set of local
deformation quantities: $U_1$, $U_1'$, $U_1''$, $U_2$, and $U_5$.

Equation~\eqref{eq:high-order-local-response-tau} has an additional structural consequence: at fixed mass and spin, $\tilde\tau^{[1]}$ is not independent of
$\tilde b^{[1]}$. The independent first-order polar information is therefore naturally represented by $(\tilde b,\tilde\gamma,\tilde\delta)$, with $\tilde\tau$ providing a consistency relation and a different observational weighting of the same local optical amplitude
$U_1$. In the adopted radial representation, a pure $U_5$ perturbation appears only in $\tilde\gamma$, while a pure $U_2$ perturbation appears only in $\tilde\delta$; the conformal function $f$ remains an exact blind direction throughout the present polar null sector.

Because the first-order observables sample only the values and a finite number of radial derivatives of $U_i(r)$ at $\tilde r_{\K}$, the polar data do not reconstruct the full radial functions. Instead, the hierarchy identifies precisely which local quantities
can be constrained, which are redundant among the four observables,
and which remain unconstrained by the present polar data. This provides the analytic basis for the metric realizations and joint-observable comparisons discussed next.

\section{Examples and deformation diagnostics}
\label{sec:examples}

We now apply the exact and weak-deviation formulas to representative metric deformations. The examples are organized by deformation channel rather than as a catalogue of alternative metrics, so that the optical origin of each leading change remains explicit. Appendix~\ref{app:metric-dictionary} gives a broader metric dictionary.

\subsection{Channel classification and representative metric realizations}
\label{subsec:examples-channel-classification}

Here and below, ``channel'' is a bookkeeping label for varying one Kerr-relative deformation function, or a specified combination of them, while the others are held at their Kerr values within the representation adopted for that mapping. The channel assignment is therefore representation dependent rather than an invariant property of the spacetime. 
The $\Delta$ and $a_1$ channels alter $\calP$ and are visible in the polar critical-curve radius and time delay. In this representation an $a_5$-only deformation leaves the critical orbit unchanged but modifies the Lyapunov exponent. A pure $a_2$ deformation leaves $(\tilde b,\tilde\gamma,\tilde\tau)$ unchanged but shifts $\tilde\delta$, whereas a pure $f$ deformation is an exact null direction of all four polar quantities.

For one-parameter comparisons, we write
\begin{equation}
\epsilon_A(r)=\chi\bar\epsilon_A(r),
\qquad
U_i(r)=\chi\bar U_i(r),
\label{eq:examples-barred-profiles}
\end{equation}
where $\chi$ is the dimensionless deformation parameter of the individual
family and the barred functions specify its unit radial profile. For the
figures below, we additionally introduce a family-dependent positive
reference scale $\chi_\star$. For families with a finite black-hole
domain, $\chi_\star$ is chosen as the boundary value of $\chi$ at the
fixed spin, corresponding to the extremal or horizon boundary as
appropriate; for the illustrative Johannsen coefficients we set
$\chi_\star=1$. The ratio $\chi/\chi_\star$ is therefore only a
normalized plotting coordinate, rather than a common measure of local
deformation strength. Equal values of $\chi/\chi_\star$, and in general
equal values of $\chi$ across distinct families, need not correspond to
equal pointwise perturbations. Table~\ref{tab:examples-unit-profiles}
lists the parameter conventions and unit radial profiles used in the
illustrations.

\begin{table*}[t]
\caption{Unit profiles for the representative one-parameter families
	used in the comparisons below. The barred quantities define the radial
	profiles per unit family parameter; multiplying them by $\chi$ gives the
	corresponding first-order deformation functions in the adopted radial
	representation.}
\label{tab:examples-unit-profiles}
\centering
\scriptsize
\renewcommand{\arraystretch}{1.12}
\begin{ruledtabular}
\begin{tabular}{p{0.15\textwidth} p{0.09\textwidth} p{0.24\textwidth} p{0.16\textwidth} p{0.15\textwidth} p{0.11\textwidth}}
Family & parameter convention & nonzero unit metric profiles & $\bar U_1$ & $\bar U_2$ & $\bar U_5$ \\ \hline
Kerr--Newman & $\chi=q^2/M^2$ & $\bar\epsilon_\Delta=M^2/\Delta_{\K}$ & $-M^2/(2\Delta_{\K})$ & $-M^2/(2\Delta_{\K})$ & $M^2/\Delta_{\K}$ \\
Bardeen & $\chi=g^2/M^2$ & $\bar\epsilon_\Delta=3M^3/(r\Delta_{\K})$ & $-3M^3/(2r\Delta_{\K})$ & $-3M^3/(2r\Delta_{\K})$ & $3M^3/(r\Delta_{\K})$ \\
Hayward & $\chi=\ell^2/M^2$ & $\bar\epsilon_\Delta=4M^4/(r^2\Delta_{\K})$ & $-2M^4/(r^2\Delta_{\K})$ & $-2M^4/(r^2\Delta_{\K})$ & $4M^4/(r^2\Delta_{\K})$ \\
Kerr--Sen & $\chi=\sigma/M$ & $\bar\epsilon_\Delta=Mr/\Delta_{\K}$, $\bar\epsilon_1=Mr/(r^2+a^2)$ & $\begin{gathered}Mr/(r^2+a^2)\\[-1pt]\quad{}-Mr/(2\Delta_{\K})\end{gathered}$ & $-Mr/(2\Delta_{\K})$ & $Mr/\Delta_{\K}$ \\
$\text{Johannsen }a_1$ & $\chi=\alpha_{13}$ & $\bar\epsilon_1=(M/r)^3$ & $(M/r)^3$ & $0$ & $0$ \\
$\text{Johannsen }a_2$ & $\chi=\alpha_{22}$ & $\bar\epsilon_2=(M/r)^2$ & $0$ & $(M/r)^2$ & $0$ \\
Simpson--Visser & $\chi=r_{\rm SV}^2/M^2$ & $\bar\epsilon_5=-(M/r)^2$ & $0$ & $0$ & $-(M/r)^2$
\end{tabular}
\end{ruledtabular}
\end{table*}

\subsection{Individual observable variations and linear regime}
\label{subsec:examples-individual-responses}

We first display the four observables separately to show the channel
selection rules and the range over which the linear approximation
remains accurate before combining them in observable space.

\begin{figure*}[t]
\centering
\IfFileExists{fig_channel_impacts_tau_delta_a0p5.pdf}{
\includegraphics[width=\textwidth]{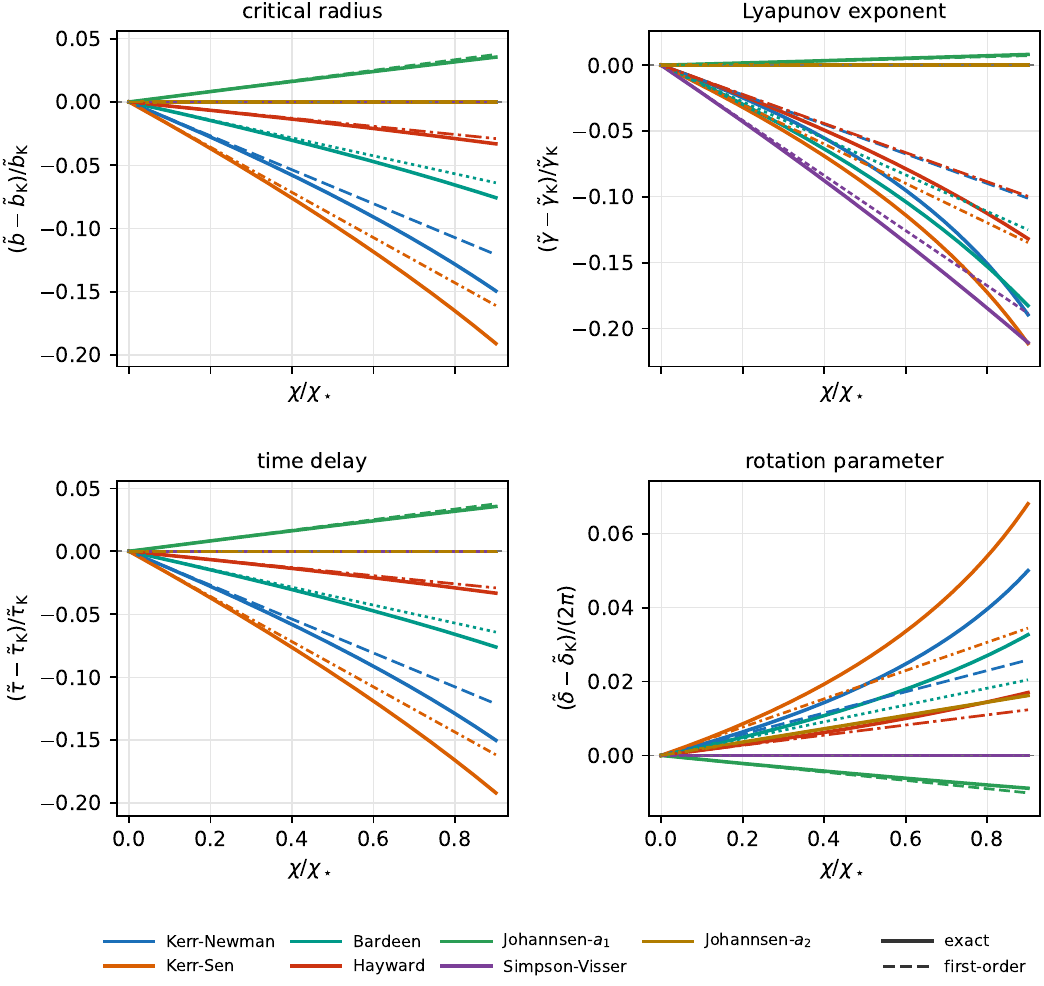}
}{
\fbox{\parbox[c][0.28\textheight][c]{0.90\textwidth}{\centering
\textit{Placeholder for \texttt{fig\_channel\_impacts\_tau\_delta\_a0p5.pdf}.}\\[3pt]
Exact and first-order variations of the four polar observables at $a/M=0.5$.}}
}
\setlength{\abovecaptionskip}{4pt}
\caption{Exact and first-order variations of the four polar observables
	at $a/M=0.5$. The panels show the fractional changes in the critical
	radius, Lyapunov exponent, and time delay, together with
	$(\tilde\delta-\tilde\delta_{\rm K})/(2\pi)$, as functions of the
	family-dependent coordinate $\chi/\chi_\star$. Solid curves are exact
	results and the model-specific patterned curves are their first-order
	approximations. Their agreement near the Kerr point illustrates the
	weak-deviation regime. The ratio $\chi/\chi_\star$ is used only as a
	common plotting coordinate and does not represent equal local deformation
	strength across different families.}
\label{fig:examples-linear-response}
\end{figure*}

Fig.~\ref{fig:examples-linear-response} shows that the first-order
curves closely track their exact counterparts near the Kerr point and
depart from them as the deformation amplitude increases. The departure from
the first-order approximation sets in at different rates for different
metric families and observables. Therefore, $\chi/\chi_\star$ should
be regarded only as a common plotting coordinate, not as a universal
measure of the validity of the linear approximation. 
The close tracking of $\tilde b$ and $\tilde\tau$ reflects both the exact
relation $\tilde\tau=2\tilde bE(a^2/\tilde b^2)$ and the linear locking in
Eq.~\eqref{eq:high-order-local-response-tau}. The isolated $a_5$ and
$a_2$ families affect only $\tilde\gamma$ and $\tilde\delta$,
respectively, while the different constant-charge and mass-function
profiles illustrate that a common channel classification does not erase
radial-profile dependence. As shown analytically below, Kerr--Newman and Kerr--Sen have identical
first-order responses when parameterized by the common amplitude $\chi$,
although their exact observables separate at finite amplitude. This
first-order degeneracy appears as a common tangent direction in
Fig.~\ref{fig:examples-observable-plane}.

\subsection{Analytic channel realizations and interpretation}
\label{subsec:examples-analytic-realizations}

\paragraph*{$\Delta$-type profiles.}
A broad class of examples has
\begin{equation}
a_1(r)=a_2(r)=a_5(r)=1,
\qquad
\Delta(r)=\Delta_{\K}(r)+h(r).
\label{eq:examples-delta-only-ansatz}
\end{equation}
The exact polar orbit obeys
\begin{equation}
4\tilde r\,\Delta(\tilde r)
-(\tilde r^2+a^2)\Delta'(\tilde r)=0,
\label{eq:examples-delta-only-orbit}
\end{equation}
and the weak optical profiles are
\begin{equation}
U_1=U_2=-\frac{h}{2\Delta_{\K}},
\qquad
U_5=\frac{h}{\Delta_{\K}}.
\label{eq:examples-delta-only-Ui}
\end{equation}
Kerr--Newman has $h=q^2$, while a signed tidal-charge continuation has $h=\beta$. These constant profiles provide the cleanest $\Delta$-channel benchmark. A separable Carter-like mass-function realization instead has
\begin{equation}
\Delta(r)=r^2-2rm(r)+a^2,
\qquad
h(r)=-2r[m(r)-M].
\label{eq:examples-mass-function-delta}
\end{equation}
The Bardeen-like and Hayward-like representatives used in Fig.~\ref{fig:examples-linear-response} are given by \cite{BambiModesto2013}
\begin{equation}
m_{\rm B}(r)=\frac{Mr^3}{(r^2+g^2)^{3/2}},
\qquad
m_{\rm H}(r)=\frac{Mr^3}{r^3+2M\ell^2}.
\label{eq:examples-bardeen-hayward-mass-functions}
\end{equation}
Their nonconstant profiles probe not only the local value of $h/\Delta_{\K}$ but also its first and second radial derivatives through $\tilde r^{[1]}$ and $\tilde\gamma^{[1]}$.

\paragraph*{Single-channel representatives and blind directions.}
For the Johannsen radial family, $\Delta=\Delta_{\K}$ and the standard asymptotically flat radial functions may be written as \cite{Johannsen2015}
\begin{align}
a_1^{\rm J}(r)&=1+\sum_{n=3}^{\infty}\alpha_{1n}\left(\frac{M}{r}\right)^n,
\nonumber\\
a_2^{\rm J}(r)&=1+\sum_{n=2}^{\infty}\alpha_{2n}\left(\frac{M}{r}\right)^n,
\nonumber\\
a_5^{\rm J}(r)&=1+\sum_{n=2}^{\infty}\alpha_{5n}\left(\frac{M}{r}\right)^n.
\label{eq:examples-johannsen-radial-series}
\end{align}
For illustration, we retain one leading coefficient at a time. 

A pure $a_1$ deformation has $\Delta=\Delta_{\K}$ and $a_1=1+\epsilon_1$, with $a_2=a_5=1$. Then
\begin{equation}
\frac{\tilde b^{[1]}}{\tilde b_{\K}}=\epsilon_1(\tilde r_{\K}),
\qquad
\tilde r^{[1]}=-\frac{\tilde b_{\K}\epsilon_1'(\tilde r_{\K})}{\calP_{\K}''(\tilde r_{\K})},
\label{eq:examples-a1-only-linear-b-and-r}
\end{equation}
The remaining responses follow from
Eqs.~\eqref{eq:high-order-local-response-gamma}--
\eqref{eq:high-order-local-response-delta}:
$\tilde\gamma^{[1]}$, $\tilde\tau^{[1]}$, and
$\tilde\delta^{[1]}$ depend respectively on
$(\epsilon_1,\epsilon_1',\epsilon_1'')$,
$\epsilon_1$, and $(\epsilon_1,\epsilon_1')$. 
For the $a_1$ example shown in the figures, we choose
\begin{equation}
a_1(r)=a_1^{\rm J}(r)=1+\alpha_{13}\left(\frac{M}{r}\right)^3.
\label{eq:examples-johannsen-a1-profile}
\end{equation}

A pure $a_2$ deformation has $\Delta=\Delta_{\K}$, $a_1=a_5=1$, and $a_2=1+\epsilon_2$. The first optical function and radial kinetic factor therefore remain exactly Kerr. Consequently,
\begin{align}
\tilde b&=\tilde b_{\K},
\qquad
\tilde\gamma=\tilde\gamma_{\K},
\qquad
\tilde\tau=\tilde\tau_{\K},
\nonumber\\
\tilde\delta-\tilde\delta_{\K}
&=2K(\mu_{\K})\calQ_{\K}(\tilde r_{\K})
\epsilon_2(\tilde r_{\K}).
\label{eq:examples-a2-only-exact}
\end{align}
For the $a_2$ example shown in the figures, we choose
\begin{equation}
a_2(r)=a_2^{\rm J}(r)=1+\epsilon_2(r)
=1+\alpha_{22}\left(\frac{M}{r}\right)^2.
\label{eq:examples-johannsen-a2-profile}
\end{equation}

A pure $a_5$ deformation in the adopted radial representation has $\Delta=\Delta_{\K}$, $a_1=a_2=1$, and $a_5=1+\epsilon_5$. Then
\begin{align}
\tilde r&=\tilde r_{\K},
\qquad
\tilde b=\tilde b_{\K},
\qquad
\tilde\tau=\tilde\tau_{\K},
\nonumber\\
\tilde\delta&=\tilde\delta_{\K},
\qquad
\frac{\tilde\gamma}{\tilde\gamma_{\K}}
=\sqrt{a_5(\tilde r_{\K})}.
\label{eq:examples-a5-only-exact}
\end{align}
For the $a_5$ example shown in the figures, we choose
\begin{equation}
a_5(r)=1-\frac{r_{\rm SV}^2}{r^2},
\label{eq:examples-SV-map}
\end{equation}
which corresponds to the rotating Simpson--Visser representation adopted here and provides an explicit realization whose polar response is instability-only in this representation  \cite{SimpsonVisser2019,Shaikh2021}. Other rotating Simpson--Visser constructions may distribute the deformation differently among the radial functions.

Finally, a pure $f$ deformation leaves the entire polar set exactly unchanged,
\begin{equation}
(\tilde r,\tilde b,\tilde\gamma,\tilde\tau,\tilde\delta)
=(\tilde r_{\K},\tilde b_{\K},\tilde\gamma_{\K},\tilde\tau_{\K},\tilde\delta_{\K}),
\label{eq:examples-exact-polar-null-direction-f}
\end{equation}
when all remaining channels take their Kerr values.

\paragraph*{Mixed realization and Kerr--Newman--Kerr--Sen tangent degeneracy.}
Kerr--Sen provides a physical mixed example \cite{Sen1992,GuoSongYan2020},
\begin{align}
f(r)&=\sigma r,
\qquad
\Delta(r)=r(r+\sigma)-2Mr+a^2,
\nonumber\\
a_1(r)&=\frac{r(r+\sigma)+a^2}{r^2+a^2},
\qquad
a_2(r)=a_5(r)=1.
\label{eq:examples-kerr-sen-map}
\end{align}
For small $\sigma$, its optical responses are
\begin{equation}
U_1=\frac{\sigma r}{r^2+a^2}-\frac{\sigma r}{2\Delta_{\K}},
\qquad
U_2=-\frac{\sigma r}{2\Delta_{\K}},
\qquad
U_5=\frac{\sigma r}{\Delta_{\K}}.
\label{eq:examples-kerr-sen-Ui}
\end{equation}
To compare the Kerr tangents of the two charge families, we adopt the common dimensionless amplitude convention
\begin{equation}
\chi\equiv
\begin{cases}
q^2/M^2, & \text{Kerr--Newman},\\
\sigma/M, & \text{Kerr--Sen}.
\end{cases}
\label{eq:examples-KN-KS-common-amplitude}
\end{equation}
The tangent comparison below is made at equal values of this common
$\chi$. Since $\chi_\star$ is family dependent, equal values of
$\chi/\chi_\star$ do not in general correspond to equal values of
$\chi$ for the two families. 
Using the Kerr polar-orbit identity
$2\tilde r_{\K}\Delta_{\K}(\tilde r_{\K})/(\tilde r_{\K}^2+a^2)=\tilde r_{\K}-M$, substitution into the response formulas of Sec.~\ref{sec:weak-deviations-from-kerr} gives the compact result
\begin{equation}
\left.
\left(
\frac{\tilde b^{[1]}}{\tilde b_{\K}},
\frac{\tilde\gamma^{[1]}}{\tilde\gamma_{\K}},
\frac{\tilde\tau^{[1]}}{\tilde\tau_{\K}},
\tilde\delta^{[1]}
\right)
\right|_{\rm KN}
=
\left.
\left(
\frac{\tilde b^{[1]}}{\tilde b_{\K}},
\frac{\tilde\gamma^{[1]}}{\tilde\gamma_{\K}},
\frac{\tilde\tau^{[1]}}{\tilde\tau_{\K}},
\tilde\delta^{[1]}
\right)
\right|_{\rm KS}.
\label{eq:examples-KN-KS-full-degeneracy}
\end{equation}
The equality holds componentwise for the size, demagnification,
time-delay, and rotation responses. The exact Kerr--Newman and Kerr--Sen observables are nevertheless distinct at finite $\chi$.  Equation~\eqref{eq:examples-KN-KS-full-degeneracy} is therefore a tangent degeneracy of the complete polar response, not an exact metric equivalence. This common tangent is directly visible in Fig.~\ref{fig:examples-observable-plane}.

\subsection{Joint observable geometry}
\label{subsec:examples-joint-response}

\begin{figure*}[t]
\centering
\IfFileExists{fig_observable_plane_tau_delta_mark02.pdf}{
\includegraphics[width=\textwidth]{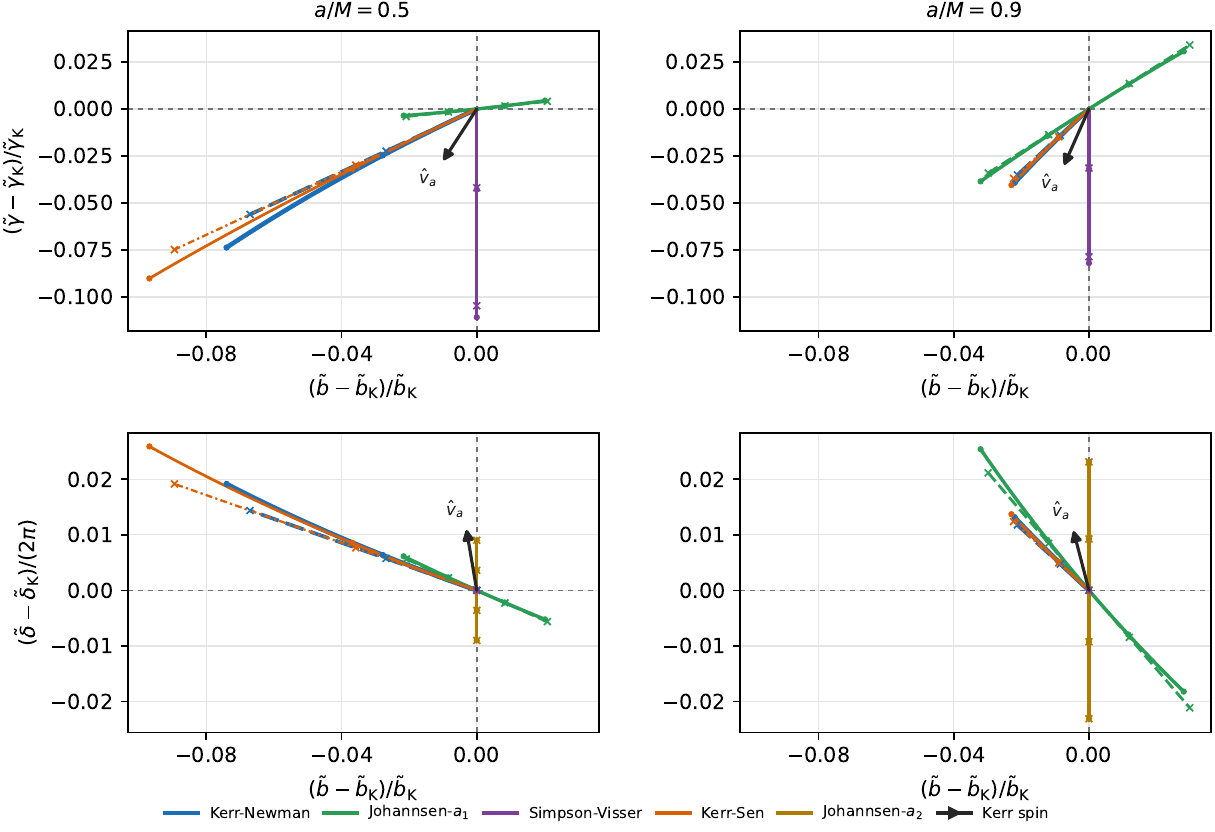}
}{
\fbox{\parbox[c][0.28\textheight][c]{0.90\textwidth}{\centering
\textit{Placeholder for \texttt{fig\_observable\_plane\_tau\_delta\_mark02.pdf}.}\\[3pt]
Joint polar-observable variations at $a/M=0.5$ and $a/M=0.9$.}}
}
\setlength{\abovecaptionskip}{4pt}
\caption{Joint variations of the polar observables for representative
	one-parameter families. The upper row shows
	$(\tilde b-\tilde b_{\rm K})/\tilde b_{\rm K}$ against
	$(\tilde\gamma-\tilde\gamma_{\rm K})/\tilde\gamma_{\rm K}$, while the
	lower row replaces the vertical coordinate by
	$(\tilde\delta-\tilde\delta_{\rm K})/(2\pi)$. Solid curves are exact
	finite-deformation trajectories and the patterned curves are their
	first-order approximations. Filled circles and crosses mark the exact and
	first-order values, respectively, at the selected amplitudes defined in
	Sec.~\ref{subsec:examples-channel-classification}. Kerr--Newman and
	Kerr--Sen share the same tangent at Kerr under the common parameter
	convention but separate at finite deformation. The black arrow denotes
	the direction of increasing Kerr spin; the opposite direction corresponds
	to decreasing spin. A Kerr mass variation at fixed $a_\ast=a/M$ is
	horizontal in both projections and is therefore not shown separately.}
\label{fig:examples-observable-plane}
\end{figure*}

The separate variations in Fig.~\ref{fig:examples-linear-response} and the hierarchy in Sec.~\ref{subsec:local-response-hierarchy} show that the four polar quantities do not define four independent directions in observable space. In particular, the first-order time-delay variation is locked to the critical-curve radius variation at fixed spin. We define
\begin{equation}
X_b=\frac{\tilde b-\tilde b_{\K}}{\tilde b_{\K}},
\qquad
X_\gamma=\frac{\tilde\gamma-\tilde\gamma_{\K}}{\tilde\gamma_{\K}},
\qquad
X_\delta=\frac{\tilde\delta-\tilde\delta_{\K}}{2\pi}.
\label{eq:examples-observable-plane-coordinates}
\end{equation}
We then use the two complementary projections $(X_b,X_\gamma)$ and $(X_b,X_\delta)$. A one-parameter non-Kerr family traces a curve in these planes, while
variation of the Kerr spin defines the Kerr tangent direction. 
A Kerr mass variation provides a second reference direction. Writing the Kerr family in terms of $(M,a_\ast)$ with $a_\ast=a/M$, a variation of $M$ at fixed $a_\ast$ scales $\tilde b_{\K}$ and $\tilde\tau_{\K}$ linearly while leaving the dimensionless $\tilde\gamma_{\K}$ and $\tilde\delta_{\K}$ unchanged. The corresponding mass tangent is therefore horizontal in both projections. No separate mass arrow is shown because this direction coincides
with the horizontal axis.

Fig.~\ref{fig:examples-observable-plane} gives a geometric summary of
the local hierarchy. The $(X_b,X_\gamma)$ projection combines
effective-size and instability information: the $\Delta$ and $a_1$
channels change both coordinates, whereas the Simpson--Visser $a_5$
realization moves vertically. The $(X_b,X_\delta)$ projection is
complementary: the $a_2$ channel is invisible to the critical-curve
radius but active in the azimuthal rotation response, while the $a_5$
realization leaves $X_\delta$ unchanged. The Bardeen-like and Hayward-like curves shown in
Fig.~\ref{fig:examples-linear-response} are omitted here for clarity;
their differences within the $\Delta$ channel are examined separately
below. 
The Kerr-spin tangent provides a direct reference for assessing local
spin--deformation degeneracies. If a non-Kerr trajectory is nearly
parallel to this tangent in a given observable plane, part of the
deformation response can be mimicked by a change in the Kerr spin.
If the deformation trajectory is not parallel to the Kerr-spin tangent,
the second coordinate carries information that cannot be reproduced by
varying the Kerr spin alone. Because these relative orientations can differ between the
$(X_b,X_\gamma)$ and $(X_b,X_\delta)$ projections, the two photon-ring
parameters provide complementary information for separating spin and
non-Kerr effects. The comparison between $a/M=0.5$ and $0.9$ further
illustrates that these relative orientations, and hence the corresponding
spin--deformation degeneracies, depend on the background Kerr spin. 

\begin{figure*}[t]
\centering
\includegraphics[width=\textwidth]{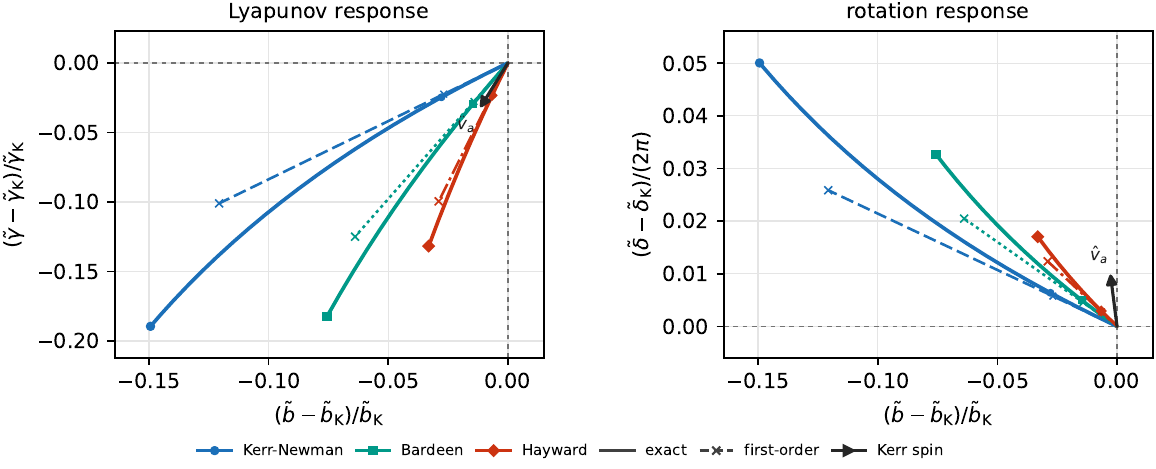}
\setlength{\abovecaptionskip}{4pt}
\caption{Joint observable variations within the $\Delta$ sector at
$a/M=0.5$. The left and right panels show the $(X_b,X_\gamma)$ and
$(X_b,X_\delta)$ projections for Kerr--Newman and the Bardeen and Hayward
mass-function profiles. The displayed deformation range and marker convention
follow Fig.~\ref{fig:examples-observable-plane}. Solid curves denote exact
finite-amplitude trajectories and the model-specific patterned curves their
first-order approximations. The black arrow gives the direction of increasing
Kerr spin. A common $\Delta$-type organization does not erase radial-profile
dependence. The $(X_b,X_\tau)$ projection is omitted because $\tilde\tau$ is
fixed by $\tilde b$ and the spin.}
\label{fig:examples-delta-channel-joint}
\end{figure*}
 
Fig.~\ref{fig:examples-delta-channel-joint} focuses on the
$\Delta$-channel sector of Fig.~\ref{fig:examples-observable-plane}.
The three $\Delta$-type realizations share the relations
$U_1=U_2=-\epsilon_\Delta/2$ and $U_5=\epsilon_\Delta$, but their distinct radial profiles lead to different local values and
derivatives entering $\tilde\gamma$ and $\tilde\delta$. Consequently,
a common channel assignment does not imply a common direction in
observable space: the photon-ring parameters retain sensitivity to the
radial shape of the deformation within the same $\Delta$ sector.
Their different alignments with the Kerr-spin direction likewise
illustrate that local spin--deformation degeneracies can depend on the
radial profile.

\section{Discussion and conclusions}
\label{sec:discussion}

We have developed a unified analytic description of the polar critical curve
and the three high-order photon-ring critical parameters within the radial
sector of a stationary, axisymmetric, separable geometry. The exact
formulation and its weak-deviation expansion show that the polar observables
probe only a restricted set of radial metric combinations and, near Kerr,
only a finite set of their local values and derivatives at the critical
orbit rather than their full radial profiles.

At the exact level, the critical orbit is the stationary point of $\calP$,
and the critical-curve radius is its stationary value. The time delay,
$\tilde\tau=2\tilde b E(a^2/\tilde b^2)$, contains no additional radial
function at fixed spin. The Lyapunov exponent additionally probes the
complete radial-instability combination
$\calW(\tilde r)\calP''(\tilde r)$, while the rotation parameter directly
samples $\calQ(\tilde r)$. The function $f$ remains an exact blind direction
of the present polar null-geodesic data set. The separation of the radial
kinetic factor from the optical curvature is representation dependent, as is
already apparent from the SBG radial function $B$; the physical statement
concerns the complete observable combination rather than $A_5$ or $\calW$
by itself \cite{SalehiWaliaChangKocherlakota2025}.

For weak deviations from Kerr, the critical-curve radius and time delay
sample the local value $U_1$, while the shift of the critical orbit samples
$U_1'$. The Lyapunov exponent depends on
$(U_1,U_1',U_1'',U_5)$, whereas the rotation parameter depends on
$(U_1,U_1',U_2)$. In particular, $\tilde\tau^{[1]}$ is locked to
$\tilde b^{[1]}$ at fixed mass and spin, so
$(\tilde b,\tilde\gamma,\tilde\delta)$ provide a minimal set of independent
first-order polar variations. Kerr--Newman and Kerr--Sen provide a
nontrivial example: under the common parameter convention in
Eq.~\eqref{eq:examples-KN-KS-common-amplitude}, all four polar variations
have the same tangent at Kerr, although their exact observables separate at
higher order.

These results follow from two complementary organizations of the same radial
separable geometry. The optical combinations $(\calP,\calQ,\calW)$ expose
the metric dependence selected by the separated null equations, while the
Kerr-relative functions $\{h,f,a_1,a_2,a_5\}$ provide a direct interface for
concrete Boyer--Lindquist-like metrics and for comparison with Kerr. A
concrete metric can therefore be inserted directly into the exact formulas
of Sec.~\ref{sec:exact-polar-observables}, and expansion of the same
functions supplies the first-order deformation profiles used in
Sec.~\ref{sec:weak-deviations-from-kerr}.

The illustrations display complementary aspects of these results. The
individual-observable comparison tests the weak-deviation approximation and
makes both the channel selection rules and the
$\tilde b$--$\tilde\tau$ relation explicit. The joint-observable planes show
how the instability and rotation information complement the critical-curve
size and how non-Kerr variations compare with changes in Kerr spin. The
$\Delta$-sector comparison further shows that a common channel classification
does not erase sensitivity to the radial deformation profile. Thus, within
the adopted representation, the deformation structure determines which
observables change at leading order, while the detailed radial profile
controls the direction and nonlinear evolution of the trajectory in
observable space. These figures are intended as geometric diagnostics rather
than a direct likelihood analysis of current horizon-scale images.

Our construction is complementary to other broad frameworks. The SBG variables $N$, $F$, $B$, and $f_{\rm SBG}$ provide a particularly transparent geometric organization of the same radial separable structure, whereas the optical and Kerr-relative representations used here are tailored to the separated calculations, direct matching of concrete metrics, and comparison with Kerr \cite{SalehiBroderickGeorgiev2024}. The Kerr-off-shell construction preserves the full Kerr hidden-symmetry tower and supplies exact general-inclination critical curves with both radial and polar deformations, while also illustrating substantial Kerr-parameter--deformation degeneracies \cite{AchourGourgoulhonRoussille2025}. The perturbative framework of Kobialko and Gal'tsov instead develops analytic general-inclination shadow morphology from a Schwarzschild-based expansion in rotation and non-Kerr deformations \cite{KobialkoGaltsov2026}. Our present emphasis is different: the exact polar critical curve radius $\tilde b$ and the full set
$(\tilde\gamma,\tilde\tau,\tilde\delta)$ are expressed in a form suitable
for direct metric substitution, and the perturbative expansion is performed about Kerr at arbitrary spin. 

At general inclination, the single polar critical-curve radius is
replaced by a full critical curve, while the three photon-ring parameters
become functions along the visible photon shell. The same analytic
organization can then be used to examine how the local polar information
identified here extends across the full photon-shell interval
\cite{SalehiWaliaChangKocherlakota2025}.

\begin{acknowledgments}
This work was supported by the National Natural Science Foundation of China under Grant No. 12305070, and the Basic Research Program of Shanxi Province under Grant Nos. 202303021222018 and 202303021221033.
\end{acknowledgments}

\appendix

\section{Metric dictionaries and relations among representations}
\label{app:metric-dictionary}

This appendix collects the relations among three complementary descriptions of the same radial separable geometry. The Johannsen functions $(A_1,A_2,A_5,f)$ specify the starting metric \cite{Johannsen2015}, the optical combinations $(\calP,\calQ,\calW)$ are selected by the separated null dynamics, and the Kerr-relative functions $(h,f,a_1,a_2,a_5)$ are defined for direct comparison with Kerr. Their exact relations are
\begin{align}
\Delta&=\Delta_{\K}+h,
&
A_1&=a_1\sqrt{\frac{\Delta_{\K}}{\Delta}},
\nonumber\\
A_2&=a_2\sqrt{\frac{\Delta_{\K}}{\Delta}},
&
A_5&=a_5\frac{\Delta}{\Delta_{\K}},
\nonumber\\
\calP&=\frac{(r^2+a^2)a_1}{\sqrt{\Delta}},
&
\calQ&=\frac{aa_2}{\sqrt{\Delta}},
\nonumber\\
\calW&=\Delta a_5.&&
\label{eq:app-representation-summary}
\end{align}

The separation of $\Delta$ from the functions $a_i$ is not unique. For any positive radial function $\Omega(r)$ that reduces to unity in
the Kerr limit, the replacement
\begin{equation}
	\begin{aligned}
		\Delta &\rightarrow \Omega^2\Delta,
		&\qquad
		a_1 &\rightarrow \Omega a_1,
		\\
		a_2 &\rightarrow \Omega a_2,
		&
		a_5 &\rightarrow \Omega^{-2}a_5
	\end{aligned}
	\label{eq:app-delta-split-freedom}
\end{equation}
leaves the original multipliers $A_i$, and therefore $(\calP,\calQ,\calW)$, unchanged. This is an algebraic freedom in the Kerr-relative representation at fixed radial coordinate, rather than a radial-coordinate transformation. The mappings below fix this freedom by specifying $\Delta$ explicitly.

The representation summarized in
Eq.~\eqref{eq:app-representation-summary} provides a convenient metric
interface: once the functions in one row of Table~\ref{tab:metric-dictionary-comprehensive} are identified, Eq.~\eqref{eq:app-representation-summary} supplies the quantities needed by the exact formulas of Sec.~\ref{sec:exact-polar-observables}. Expanding the same functions about their Kerr values then gives the $U_i$ profiles used in Sec.~\ref{sec:weak-deviations-from-kerr}. The table therefore records mappings rather than independent metric ans\"atze.
For the Johannsen radial family, $h=0$ and hence $a_i^{\rm J}=A_i^{\rm J}$; the standard series for $a_i^{\rm J}$ are given in Eq.~\eqref{eq:examples-johannsen-radial-series}, while $f_{\rm J}(r)=\sum_{n=3}^{\infty}\epsilon_n M^n/r^{n-2}$ in the usual Johannsen convention \cite{Johannsen2015}. In the Baines--Visser row, $\Delta_{\rm BV}(r)$, $\Xi(r)$, and $\Phi(r)$ denote the three arbitrary radial functions of that representation \cite{BainesVisser2023}. In the Azreg--A\"inou row, $\mathfrak f_{\rm A}(r)$ is the arbitrary radial function entering $\Delta_{\rm A}=r^2+a^2-2\mathfrak f_{\rm A}(r)$ \cite{AzregAinou2014}. 
The mappings use the conventions of Refs.~\cite{AlievGumrukcuoglu2005,Moffat2015,GuoObersYan2018,BambiModesto2013,Sen1992,SalehiBroderickGeorgiev2024,BainesVisser2023,AzregAinou2014,SimpsonVisser2019,Shaikh2021}. In the Kerr--MOG row, $M$ denotes the asymptotic mass. As throughout the main text, the radial coordinate is the Boyer--Lindquist-like coordinate used in the corresponding mapping.

\begin{table*}[t]
\caption{Radial-function dictionary for representative separable metrics and
	metric families. Each row specifies the adopted Kerr-relative mapping,
	with $\Delta(r)=\Delta_{\K}(r)+h(r)$. The listed functions can be inserted
	directly into the exact formulas of Sec.~\ref{sec:exact-polar-observables};
	expanding the same functions about their Kerr values gives the first-order
	deformation profiles used in Sec.~\ref{sec:weak-deviations-from-kerr}. The
	Johannsen, Baines--Visser, and Azreg--A\"inou functions are defined in the
	preceding discussion.}
\label{tab:metric-dictionary-comprehensive}
\centering
\scriptsize
\setlength{\tabcolsep}{2.2pt}
\renewcommand{\arraystretch}{1.30}
\begin{ruledtabular}
\begin{tabular}{p{0.180\textwidth} p{0.300\textwidth} p{0.095\textwidth} p{0.145\textwidth} p{0.055\textwidth} p{0.105\textwidth}}
Metric or family & $h(r)$ & $f(r)$ & $a_1(r)$ & $a_2(r)$ & $a_5(r)$ \\
\hline
Kerr
& $0$
& $0$
& $1$ & $1$ & $1$ \\
Kerr--Newman
& $q^2$
& $0$
& $1$ & $1$ & $1$ \\
Tidal charge
& $\beta$
& $0$
& $1$ & $1$ & $1$ \\
Kerr--MOG
& $\dfrac{\alpha}{1+\alpha}M^2$
& $0$
& $1$ & $1$ & $1$ \\
Bardeen
& $\displaystyle 2Mr\!\left[1-\frac{r^3}{(r^2+g^2)^{3/2}}\right]$
& $0$
& $1$ & $1$ & $1$ \\
Hayward
& $\displaystyle \frac{4M^2\ell^2 r}{r^3+2M\ell^2}$
& $0$
& $1$ & $1$ & $1$ \\
Ay\'on--Beato--Garc\'ia
& $\displaystyle \begin{gathered}
2Mr\!\left[1-\dfrac{r^3}{(r^2+Q^2)^{3/2}}\right]\\[-1pt]
{}+\dfrac{Q^2r^4}{(r^2+Q^2)^2}
\end{gathered}$
& $0$
& $1$ & $1$ & $1$ \\
Kerr--Sen
& $\sigma r$
& $\sigma r$
& $\displaystyle 1+\frac{\sigma r}{r^2+a^2}$
& $1$ & $1$ \\
Simpson--Visser
& $0$
& $0$
& $1$ & $1$
& $\displaystyle 1-\frac{r_{\rm SV}^2}{r^2}$ \\
Johannsen radial family
& $0$
& $f_{\rm J}(r)$
& $a_1^{\rm J}(r)$
& $a_2^{\rm J}(r)$
& $a_5^{\rm J}(r)$ \\
Baines--Visser family
& $\displaystyle \begin{gathered}
\Delta_{\rm BV}(r)e^{-2\Phi(r)}
-\Delta_{\K}(r)
\end{gathered}$
& $\Xi^2(r)-r^2$
& $\displaystyle \frac{\Xi^2(r)+a^2}{r^2+a^2}$
& $1$
& $e^{2\Phi(r)}$ \\
Azreg--A\"inou family
& $2\!\left[Mr-\mathfrak f_{\rm A}(r)\right]$
& $0$
& $1$ & $1$ & $1$ \\
\end{tabular}
\end{ruledtabular}
\end{table*}

\paragraph*{Relation to the SBG representation.}
The SBG variables \cite{SalehiBroderickGeorgiev2024} are connected directly to $(\calP,\calQ,\calW)$ by
\begin{equation}
	\calP=\frac{r}{N},\qquad
	\calQ=\frac{aF}{N},\qquad
	\calW=\frac{r^2N^2}{B^2},
	\qquad
	f=f_{\rm SBG}.
	\label{eq:app-SBG-optical-map}
\end{equation}
Motivated by the radial combination $N^2/F^2$ used in the SBG representation, we choose
\begin{equation}
	\Delta=\frac{N^2}{F^2}.
	\label{eq:app-SBG-delta-choice}
\end{equation}
With the standard exterior branch $\sqrt{\Delta}=N/F$, Eq.~\eqref{eq:app-representation-summary} then gives
\begin{equation}
	\begin{aligned}
		h&=\frac{N^2}{F^2}-\Delta_{\K},
		&\qquad
		a_1&=\frac{r}{F(r^2+a^2)},
		\\
		a_2&=1,
		&
		a_5&=\frac{r^2F^2}{B^2}.
	\end{aligned}
	\label{eq:metric-dictionary-SBG-map}
\end{equation}
Thus $a_2=1$ follows from this choice and does not restrict the general $a_2(r)$ sector. Within the same mapping, varying $a_5$ at fixed $N$ and $F$ corresponds to varying the radial-normalization function $B$. The polar critical condition $\calP'(\tilde r)=0$ then reduces directly to the SBG relation obtained from $r/N$. This comparison also illustrates the complementary roles of the two descriptions: $(N,F,B,f)$ emphasizes geometrical sectors, whereas $(\calP,\calQ,\calW)$ and $(h,f,a_i)$ expose the combinations needed for the null-geodesic calculation and for comparison with Kerr.

\section{Useful exact and Kerr-reference formulas}
\label{app:useful-formulas}

For practical use of the exact and first-order results, we collect here
several radial derivatives that would otherwise interrupt the main
derivation.

\paragraph*{Exact optical derivatives.}
From Eq.~\eqref{eq:metric-kerr-adapted-optical-functions},
\begin{align}
\calP'(r)
={}&
\frac{2ra_1+(r^2+a^2)a_1'}{\sqrt{\Delta}}
-\frac{(r^2+a^2)a_1\Delta'}{2\Delta^{3/2}},
\label{eq:app-calP-first-derivative}
\\
\calQ'(r)
={}&
\frac{aa_2'}{\sqrt{\Delta}}
-\frac{aa_2\Delta'}{2\Delta^{3/2}},
\label{eq:app-calQ-first-derivative}
\end{align}
and
\begin{align}
\calP''(r)
={}&
\frac{2a_1+4ra_1'+(r^2+a^2)a_1''}{\sqrt{\Delta}}
\nonumber\\
&-\frac{\Delta'}{\Delta^{3/2}}
\left[2ra_1+(r^2+a^2)a_1'\right]
\nonumber\\
&-\frac{\Delta''(r^2+a^2)a_1}{2\Delta^{3/2}}
+\frac{3\Delta'^2(r^2+a^2)a_1}{4\Delta^{5/2}}.
\label{eq:high-order-rings-calP-second-derivative}
\end{align}
All functions on the right-hand sides are evaluated at the same radial argument before setting $r=\tilde r$.

\paragraph*{Kerr radial derivatives.}
For the Kerr factors used in Sec.~\ref{sec:weak-deviations-from-kerr},
\begin{equation}
\calP_{\K}'(r)
=
\frac{r^3-3Mr^2+a^2r+Ma^2}{\Delta_{\K}^{3/2}},
\label{eq:app-Kerr-calP-first}
\end{equation}
so that its numerator gives Eq.~\eqref{eq:polar-kerr-critical-radius-cubic}. The derivatives entering the first-order coefficients can be written as
\begin{align}
\calP_{\K}''(r)
={}&
\frac{
3M^2(r^2+a^2)-8Ma^2r+a^2(r^2+a^2)
}{\Delta_{\K}^{5/2}},
\label{eq:app-Kerr-calP-second}
\\
\frac{\calP_{\K}'''(r)}{\calP_{\K}''(r)}
={}&
\frac{2r(3M^2+a^2)-8Ma^2}
{3M^2(r^2+a^2)-8Ma^2r+a^2(r^2+a^2)}
\nonumber\\
&-5\frac{r-M}{\Delta_{\K}(r)},
\label{eq:app-Kerr-calP-third-ratio}
\\
\calQ_{\K}'(r)
={}&
-\frac{a(r-M)}{\Delta_{\K}^{3/2}},
\qquad
\frac{\Delta_{\K}'(r)}{\Delta_{\K}(r)}
=
\frac{2(r-M)}{\Delta_{\K}(r)}.
\label{eq:app-Kerr-Q-Delta-derivatives}
\end{align}
Evaluating Eqs.~\eqref{eq:app-Kerr-calP-second}--\eqref{eq:app-Kerr-Q-Delta-derivatives} at $r=\tilde r_{\K}$ gives the coefficients appearing in Eqs.~\eqref{eq:high-order-rings-kappa-response} and \eqref{eq:high-order-local-response-gamma}. Further algebraic reductions using Eq.~\eqref{eq:polar-kerr-critical-radius-cubic} are possible, but the forms above keep the dependence on $(M,a,\tilde r_{\K})$ explicit without introducing additional auxiliary notation.

\bibliography{references}

@article{Carter1968,
  author  = {Carter, Brandon},
  title   = {Global Structure of the {Kerr} Family of Gravitational Fields},
  journal = {Phys. Rev.},
  volume  = {174},
  pages   = {1559--1571},
  year    = {1968},
  doi     = {10.1103/PhysRev.174.1559}
}

@incollection{Bardeen1973,
  author    = {Bardeen, James M.},
  title     = {Timelike and Null Geodesics in the {Kerr} Metric},
  booktitle = {Black Holes (Les Astres Occlus)},
  editor    = {DeWitt, C. and DeWitt, B. S.},
  publisher = {Gordon and Breach},
  address   = {New York},
  pages     = {215--239},
  year      = {1973}
}

@article{GrallaLupsasca2020,
  author        = {Gralla, Samuel E. and Lupsasca, Alexandru},
  title         = {Lensing by {Kerr} Black Holes},
  journal       = {Phys. Rev. D},
  volume        = {101},
  pages         = {044031},
  year          = {2020},
  doi           = {10.1103/PhysRevD.101.044031},
  eprint        = {1910.12873},
  archivePrefix = {arXiv},
  primaryClass  = {gr-qc}
}

@article{Johnson2020,
  author        = {Johnson, Michael D. and Lupsasca, Alexandru and Strominger, Andrew and Wong, George N. and Hadar, Shahar and Kapec, Daniel and Narayan, Ramesh and Chael, Andrew and Gammie, Charles F. and Galison, Peter and Palumbo, Daniel C. M. and Doeleman, Sheperd S. and Blackburn, Lindy and Wielgus, Maciek and Pesce, Dominic W. and Farah, Joseph R. and Moran, James M.},
  title         = {Universal Interferometric Signatures of a Black Hole's Photon Ring},
  journal       = {Sci. Adv.},
  volume        = {6},
  pages         = {eaaz1310},
  year          = {2020},
  doi           = {10.1126/sciadv.aaz1310},
  eprint        = {1907.04329},
  archivePrefix = {arXiv},
  primaryClass  = {astro-ph.IM}
}

@article{GrallaHolzWald2019,
  author        = {Gralla, Samuel E. and Holz, Daniel E. and Wald, Robert M.},
  title         = {Black Hole Shadows, Photon Rings, and Lensing Rings},
  journal       = {Phys. Rev. D},
  volume        = {100},
  pages         = {024018},
  year          = {2019},
  doi           = {10.1103/PhysRevD.100.024018},
  eprint        = {1906.00873},
  archivePrefix = {arXiv},
  primaryClass  = {astro-ph.HE}
}

@article{EHTM87,
  author        = {{Event Horizon Telescope Collaboration}},
  title         = {First {M87} {Event Horizon Telescope} Results. {I}. The Shadow of the Supermassive Black Hole},
  journal       = {Astrophys. J. Lett.},
  volume        = {875},
  number        = {1},
  pages         = {L1},
  year          = {2019},
  doi           = {10.3847/2041-8213/ab0ec7},
  eprint        = {1906.11238},
  archivePrefix = {arXiv},
  primaryClass  = {astro-ph.GA}
}

@article{EHTSgrA,
  author        = {{Event Horizon Telescope Collaboration}},
  title         = {First {Sagittarius A*} {Event Horizon Telescope} Results. {I}. The Shadow of the Supermassive Black Hole in the Center of the {Milky Way}},
  journal       = {Astrophys. J. Lett.},
  volume        = {930},
  number        = {2},
  pages         = {L12},
  year          = {2022},
  doi           = {10.3847/2041-8213/ac6674},
  eprint        = {2311.08680},
  archivePrefix = {arXiv},
  primaryClass  = {astro-ph.GA}
}

@article{Johannsen2015,
  author        = {Johannsen, Tim},
  title         = {Regular Black Hole Metric with Three Constants of Motion},
  journal       = {Phys. Rev. D},
  volume        = {88},
  pages         = {044002},
  year          = {2013},
  doi           = {10.1103/PhysRevD.88.044002},
  eprint        = {1501.02809},
  archivePrefix = {arXiv},
  primaryClass  = {gr-qc}
}

@article{JohannsenPhotonRings2013,
  author        = {Johannsen, Tim},
  title         = {Photon Rings around {Kerr} and {Kerr-like} Black Holes},
  journal       = {Astrophys. J.},
  volume        = {777},
  pages         = {170},
  year          = {2013},
  doi           = {10.1088/0004-637X/777/2/170},
  eprint        = {1501.02814},
  archivePrefix = {arXiv},
  primaryClass  = {astro-ph.HE}
}

@article{SalehiBroderickGeorgiev2024,
  author        = {Salehi, Kiana and Broderick, Avery E. and Georgiev, Boris},
  title         = {Photon Rings and Shadow Size for General Axisymmetric and Stationary Integrable Spacetimes},
  journal       = {Astrophys. J.},
  volume        = {966},
  pages         = {143},
  year          = {2024},
  doi           = {10.3847/1538-4357/ad37fa},
  eprint        = {2311.01495},
  archivePrefix = {arXiv},
  primaryClass  = {gr-qc}
}

@article{SalehiWaliaChangKocherlakota2025,
  author        = {Salehi, Kiana and Walia, Rahul Kumar and Chang, Dominic O. and Kocherlakota, Prashant},
  title         = {Influence of Observer's Inclination and Spacetime Structure on Photon Ring Observables},
  journal       = {Phys. Rev. D},
  volume        = {111},
  pages         = {104057},
  year          = {2025},
  doi           = {10.1103/PhysRevD.111.104057},
  eprint        = {2411.15310},
  archivePrefix = {arXiv},
  primaryClass  = {gr-qc}
}

@article{WaliaKocherlakotaChangSalehi2025,
  author        = {Walia, Rahul Kumar and Kocherlakota, Prashant and Chang, Dominic O. and Salehi, Kiana},
  title         = {Spacetime Measurements with the Photon Ring},
  journal       = {Phys. Rev. D},
  volume        = {111},
  pages         = {104074},
  year          = {2025},
  doi           = {10.1103/PhysRevD.111.104074},
  eprint        = {2411.15119},
  archivePrefix = {arXiv},
  primaryClass  = {gr-qc}
}

@article{BainesVisser2023,
  author        = {Baines, Joshua and Visser, Matt},
  title         = {Killing Horizons and Surface Gravities for a Well-Behaved Three-Function Generalization of the {Kerr} Spacetime},
  journal       = {Universe},
  volume        = {9},
  pages         = {223},
  year          = {2023},
  doi           = {10.3390/universe9050223},
  eprint        = {2303.07380},
  archivePrefix = {arXiv},
  primaryClass  = {gr-qc}
}

@article{AzregAinou2014,
  author        = {Azreg-Ainou, Mustapha},
  title         = {Generating Rotating Regular Black Hole Solutions without Complexification},
  journal       = {Phys. Rev. D},
  volume        = {90},
  pages         = {064041},
  year          = {2014},
  doi           = {10.1103/PhysRevD.90.064041},
  eprint        = {1405.2569},
  archivePrefix = {arXiv},
  primaryClass  = {gr-qc}
}

@article{AchourGourgoulhonRoussille2025,
  author        = {Ben Achour, Jibril and Gourgoulhon, Eric and Roussille, Hugo},
  title         = {Black Hole Photon Ring beyond General Relativity: An Integrable Parametrization},
  journal       = {JCAP},
  volume        = {2025},
  number        = {10},
  pages         = {012},
  year          = {2025},
  doi           = {10.1088/1475-7516/2025/10/012},
  eprint        = {2506.09882},
  archivePrefix = {arXiv},
  primaryClass  = {gr-qc}
}

@article{KobialkoGaltsov2026,
  author        = {Kobialko, Kirill and Gal'tsov, Dmitri},
  title         = {Perturbation Theory for Gravitational Shadows in {Kerr-like} Spacetimes},
  journal       = {Phys. Rev. D},
  volume        = {113},
  pages         = {104007},
  year          = {2026},
  doi           = {10.1103/7fkt-1knw},
  eprint        = {2512.24126},
  archivePrefix = {arXiv},
  primaryClass  = {gr-qc}
}

@article{DeichYunesGammie2024,
  author        = {Deich, Alexander and Yunes, Nicolas and Gammie, Charles F.},
  title         = {Lyapunov Exponents to Test General Relativity},
  journal       = {Phys. Rev. D},
  volume        = {110},
  pages         = {044033},
  year          = {2024},
  doi           = {10.1103/PhysRevD.110.044033},
  eprint        = {2308.07232},
  archivePrefix = {arXiv},
  primaryClass  = {gr-qc}
}

@article{Sen1992,
  author        = {Sen, Ashoke},
  title         = {Rotating Charged Black Hole Solution in Heterotic String Theory},
  journal       = {Phys. Rev. Lett.},
  volume        = {69},
  pages         = {1006--1009},
  year          = {1992},
  doi           = {10.1103/PhysRevLett.69.1006},
  eprint        = {hep-th/9204046},
  archivePrefix = {arXiv}
}

@article{AlievGumrukcuoglu2005,
  author        = {Aliev, A. N. and Gumrukcuoglu, A. E.},
  title         = {Charged Rotating Black Holes on a 3-Brane},
  journal       = {Phys. Rev. D},
  volume        = {71},
  pages         = {104027},
  year          = {2005},
  doi           = {10.1103/PhysRevD.71.104027},
  eprint        = {hep-th/0502223},
  archivePrefix = {arXiv}
}

@article{BambiModesto2013,
  author        = {Bambi, Cosimo and Modesto, Leonardo},
  title         = {Rotating Regular Black Holes},
  journal       = {Phys. Lett. B},
  volume        = {721},
  pages         = {329--334},
  year          = {2013},
  doi           = {10.1016/j.physletb.2013.03.025},
  eprint        = {1302.6075},
  archivePrefix = {arXiv},
  primaryClass  = {gr-qc}
}

@article{SimpsonVisser2019,
  author        = {Simpson, Alex and Visser, Matt},
  title         = {Black-Bounce to Traversable Wormhole},
  journal       = {JCAP},
  volume        = {2019},
  number        = {02},
  pages         = {042},
  year          = {2019},
  doi           = {10.1088/1475-7516/2019/02/042},
  eprint        = {1812.07114},
  archivePrefix = {arXiv},
  primaryClass  = {gr-qc}
}

@article{Shaikh2021,
  author        = {Shaikh, Rajibul and Pal, Kunal and Pal, Kuntal and Sarkar, Tapobrata},
  title         = {Constraining Alternatives to the {Kerr} Black Hole},
  journal       = {Mon. Not. R. Astron. Soc.},
  volume        = {506},
  pages         = {1229--1236},
  year          = {2021},
  doi           = {10.1093/mnras/stab1779},
  eprint        = {2102.04299},
  archivePrefix = {arXiv},
  primaryClass  = {gr-qc}
}

@article{GrallaLupsascaObservable2020,
  author        = {Gralla, Samuel E. and Lupsasca, Alexandru},
  title         = {On the Observable Shape of Black Hole Photon Rings},
  journal       = {Phys. Rev. D},
  volume        = {102},
  pages         = {124003},
  year          = {2020},
  doi           = {10.1103/PhysRevD.102.124003},
  eprint        = {2007.10336},
  archivePrefix = {arXiv},
  primaryClass  = {gr-qc}
}

@article{GrallaLupsascaMarrone2020,
  author        = {Gralla, Samuel E. and Lupsasca, Alexandru and Marrone, Daniel P.},
  title         = {The Shape of the Black Hole Photon Ring: A Precise Test of Strong-Field General Relativity},
  journal       = {Phys. Rev. D},
  volume        = {102},
  pages         = {124004},
  year          = {2020},
  doi           = {10.1103/PhysRevD.102.124004},
  eprint        = {2008.03879},
  archivePrefix = {arXiv},
  primaryClass  = {gr-qc}
}

@article{StaelensMayersonBacchiniRipperdaKuchler2023,
  author        = {Staelens, Seppe and Mayerson, Daniel R. and Bacchini, Fabio and Ripperda, Bart and K{\"u}chler, Lorenzo},
  title         = {Black Hole Photon Rings Beyond General Relativity},
  journal       = {Phys. Rev. D},
  volume        = {107},
  pages         = {124026},
  year          = {2023},
  doi           = {10.1103/PhysRevD.107.124026},
  eprint        = {2303.02111},
  archivePrefix = {arXiv},
  primaryClass  = {gr-qc}
}

@article{CarsonYagi2020,
  author        = {Carson, Zack and Yagi, Kent},
  title         = {Asymptotically Flat, Parameterized Black Hole Metric Preserving {Kerr} Symmetries},
  journal       = {Phys. Rev. D},
  volume        = {101},
  pages         = {084030},
  year          = {2020},
  doi           = {10.1103/PhysRevD.101.084030},
  eprint        = {2002.01028},
  archivePrefix = {arXiv},
  primaryClass  = {gr-qc}
}

@article{GlampedakisPappas2019,
  author        = {Glampedakis, Kostas and Pappas, George},
  title         = {The Modification of Photon Trapping Orbits as a Diagnostic of Non-{Kerr} Spacetimes},
  journal       = {Phys. Rev. D},
  volume        = {99},
  pages         = {124041},
  year          = {2019},
  doi           = {10.1103/PhysRevD.99.124041},
  eprint        = {1806.09333},
  archivePrefix = {arXiv},
  primaryClass  = {gr-qc}
}

@article{HadarJohnsonLupsascaWong2021,
  author        = {Hadar, Shahar and Johnson, Michael D. and Lupsasca, Alexandru and Wong, George N.},
  title         = {Photon Ring Autocorrelations},
  journal       = {Phys. Rev. D},
  volume        = {103},
  pages         = {104038},
  year          = {2021},
  doi           = {10.1103/PhysRevD.103.104038},
  eprint        = {2010.03683},
  archivePrefix = {arXiv},
  primaryClass  = {gr-qc}
}

@article{Moffat2015,
  author        = {Moffat, J. W.},
  title         = {Black Holes in Modified Gravity ({MOG})},
  journal       = {Eur. Phys. J. C},
  volume        = {75},
  pages         = {175},
  year          = {2015},
  doi           = {10.1140/epjc/s10052-015-3405-x},
  eprint        = {1412.5424},
  archivePrefix = {arXiv},
  primaryClass  = {gr-qc}
}

@article{PapadopoulosKokkotas2018,
  author        = {Papadopoulos, Georgios O. and Kokkotas, Kostas D.},
  title         = {Preserving {Kerr} Symmetries in Deformed Spacetimes},
  journal       = {Class. Quant. Grav.},
  volume        = {35},
  pages         = {185014},
  year          = {2018},
  doi           = {10.1088/1361-6382/aad7f4},
  eprint        = {1807.08594},
  archivePrefix = {arXiv},
  primaryClass  = {gr-qc}
}

@article{PaugnatLupsascaVincentWielgus2022,
  author        = {Paugnat, Hadrien and Lupsasca, Alexandru and Vincent, Fr{\'e}d{\'e}ric H. and Wielgus, Maciek},
  title         = {Photon Ring Test of the {Kerr} Hypothesis: Variation in the Ring Shape},
  journal       = {Astron. Astrophys.},
  volume        = {668},
  pages         = {A11},
  year          = {2022},
  doi           = {10.1051/0004-6361/202244216},
  eprint        = {2206.02781},
  archivePrefix = {arXiv},
  primaryClass  = {astro-ph.HE}
}

@article{PerlickTsupko2022,
  author        = {Perlick, Volker and Tsupko, Oleg Yu.},
  title         = {Calculating Black Hole Shadows: Review of Analytical Studies},
  journal       = {Phys. Rept.},
  volume        = {947},
  pages         = {1--39},
  year          = {2022},
  doi           = {10.1016/j.physrep.2021.10.004},
  eprint        = {2105.07101},
  archivePrefix = {arXiv},
  primaryClass  = {gr-qc}
}

@article{KonoplyaRezzollaZhidenko2016,
  author        = {Konoplya, Roman and Rezzolla, Luciano and Zhidenko, Alexander},
  title         = {General Parametrization of Axisymmetric Black Holes in Metric Theories of Gravity},
  journal       = {Phys. Rev. D},
  volume        = {93},
  pages         = {064015},
  year          = {2016},
  doi           = {10.1103/PhysRevD.93.064015},
  eprint        = {1602.02378},
  archivePrefix = {arXiv},
  primaryClass  = {gr-qc}
}

@article{CardosoPaniRico2014,
  author        = {Cardoso, Vitor and Pani, Paolo and Rico, Jo{\~a}o},
  title         = {On Generic Parametrizations of Spinning Black-Hole Geometries},
  journal       = {Phys. Rev. D},
  volume        = {89},
  pages         = {064007},
  year          = {2014},
  doi           = {10.1103/PhysRevD.89.064007},
  eprint        = {1401.0528},
  archivePrefix = {arXiv},
  primaryClass  = {gr-qc}
}

@article{Teo2003,
  author  = {Teo, Edward},
  title   = {Spherical Photon Orbits Around a {Kerr} Black Hole},
  journal = {Gen. Rel. Grav.},
  volume  = {35},
  pages   = {1909--1926},
  year    = {2003},
  doi     = {10.1023/A:1026286607562}
}

@article{GiataganasKehagiasRiotto2024,
  author        = {Giataganas, D. and Kehagias, A. and Riotto, A.},
  title         = {Quasinormal Modes and Universality of the {Penrose} Limit of Black Hole Photon Rings},
  journal       = {JHEP},
  volume        = {2024},
  number        = {09},
  pages         = {168},
  year          = {2024},
  doi           = {10.1007/JHEP09(2024)168},
  eprint        = {2403.10605},
  archivePrefix = {arXiv},
  primaryClass  = {gr-qc}
}

@article{HouLiuGuoYanChen2022,
  author        = {Hou, Yehui and Liu, Peng and Guo, Minyong and Yan, Haopeng and Chen, Bin},
  title         = {Multi-level Images Around {Kerr--Newman} Black Holes},
  journal       = {Class. Quant. Grav.},
  volume        = {39},
  number        = {19},
  pages         = {194001},
  year          = {2022},
  doi           = {10.1088/1361-6382/ac8860},
  eprint        = {2203.02755},
  archivePrefix = {arXiv},
  primaryClass  = {gr-qc}
}

@article{GuoSongYan2020,
  author        = {Guo, Minyong and Song, Shupeng and Yan, Haopeng},
  title         = {Observational Signature of a Near-Extremal {Kerr--Sen} Black Hole in the Heterotic String Theory},
  journal       = {Phys. Rev. D},
  volume        = {101},
  number        = {2},
  pages         = {024055},
  year          = {2020},
  doi           = {10.1103/PhysRevD.101.024055},
  eprint        = {1911.04796},
  archivePrefix = {arXiv},
  primaryClass  = {gr-qc}
}

@article{GuoObersYan2018,
  author        = {Guo, Minyong and Obers, Niels A. and Yan, Haopeng},
  title         = {Observational Signatures of Near-Extremal {Kerr-like} Black Holes in a Modified Gravity Theory at the {Event Horizon Telescope}},
  journal       = {Phys. Rev. D},
  volume        = {98},
  number        = {8},
  pages         = {084063},
  year          = {2018},
  doi           = {10.1103/PhysRevD.98.084063},
  eprint        = {1806.05249},
  archivePrefix = {arXiv},
  primaryClass  = {gr-qc}
}

@article{GuoGao2021,
  author        = {Guo, Minyong and Gao, Sijie},
  title         = {Universal Properties of Light Rings for Stationary Axisymmetric Spacetimes},
  journal       = {Phys. Rev. D},
  volume        = {103},
  number        = {10},
  pages         = {104031},
  year          = {2021},
  doi           = {10.1103/PhysRevD.103.104031},
  eprint        = {2011.02211},
  archivePrefix = {arXiv},
  primaryClass  = {gr-qc}
}

@article{GuoZhongWangGao2022,
  author        = {Guo, Minyong and Zhong, Zhen and Wang, Jinguang and Gao, Sijie},
  title         = {Light Rings and Long-Lived Modes in Quasiblack Hole Spacetimes},
  journal       = {Phys. Rev. D},
  volume        = {105},
  number        = {2},
  pages         = {024049},
  year          = {2022},
  doi           = {10.1103/PhysRevD.105.024049},
  eprint        = {2108.08967},
  archivePrefix = {arXiv},
  primaryClass  = {gr-qc}
}

@article{WeiLiu2021,
	author        = {Shao-Wen Wei and Yu-Xiao Liu},
	title         = {Testing the nature of {Gauss--Bonnet} gravity by
	four-dimensional rotating black hole shadow},
	journal       = {Eur. Phys. J. Plus},
	volume        = {136},
	pages         = {436},
	year          = {2021},
	doi           = {10.1140/epjp/s13360-021-01398-9},
	eprint        = {2003.07769},
	archivePrefix = {arXiv},
	primaryClass  = {gr-qc}
}

@article{SunLiuQianChenYue2023,
	author  = {Jiaojiao Sun and Yunqi Liu and Wei-Liang Qian
	and Songbai Chen and Ruihong Yue},
	title   = {Black hole shadow in {$f(R)$} gravity with nonlinear electrodynamics},
	journal = {Chinese Phys. C},
	volume  = {47},
	pages   = {025104},
	year    = {2023},
	doi     = {10.1088/1674-1137/aca4bc}
}

@article{ZhengWuLiJiang2025,
	author        = {He-Bin Zheng and Meng-Qi Wu and Guo-Ping Li and Qing-Quan Jiang},
	title         = {Shadows and accretion disk images of charged rotating black hole
	in modified gravity theory},
	journal       = {Eur. Phys. J. C},
	volume        = {85},
	pages         = {46},
	year          = {2025},
	doi           = {10.1140/epjc/s10052-025-13791-0},
	eprint        = {2411.10315},
	archivePrefix = {arXiv},
	primaryClass  = {gr-qc}
}

@article{LiZhengHeJiang2025,
	author        = {Guo-Ping Li and He-Bin Zheng and Ke-Jian He and Qing-Quan Jiang},
	title         = {The shadow and observational images of the non-singular
	rotating black holes in loop quantum gravity},
	journal       = {Eur. Phys. J. C},
	volume        = {85},
	pages         = {249},
	year          = {2025},
	doi           = {10.1140/epjc/s10052-025-13997-2},
	eprint        = {2410.17295},
	archivePrefix = {arXiv},
	primaryClass  = {gr-qc}
}

@article{HeYangZengChang2026,
	author  = {Ke-Jian He and Chen-Yu Yang and Xiao-Xiong Zeng and Zheng-Xue Chang},
	title   = {Shadow images of a rotating black hole in {Kalb--Ramond} gravity
	surrounded by the thin accretion disk},
	journal = {Chinese Phys. C},
	volume  = {50},
	pages   = {035102},
	year    = {2026},
	doi     = {10.1088/1674-1137/ae1aff}
}

@article{YangYeZeng2026,
	author        = {Chen-Yu Yang and Huan Ye and Xiao-Xiong Zeng},
	title         = {Shadow and polarization images of rotating black holes in
	{Kalb--Ramond} gravity illuminated by several thick accretion disks},
	journal       = {Eur. Phys. J. C},
	volume        = {86},
	pages         = {370},
	year          = {2026},
	doi           = {10.1140/epjc/s10052-026-15584-5},
	eprint        = {2510.21229},
	archivePrefix = {arXiv},
	primaryClass  = {gr-qc}
}

@misc{WangWangWangGuo2026,
	author        = {Xinyu Wang and Xiaobao Wang and Xin-Yang Wang and Minyong Guo},
	title         = {Novel extended inner shadow in images of {Johannsen--Psaltis}
	black holes with thin accretion disks},
	year          = {2026},
	eprint        = {2607.18751},
	archivePrefix = {arXiv},
	primaryClass  = {gr-qc}
}

@misc{LongChenJing2026,
	author        = {Fen Long and Songbai Chen and Jiliang Jing},
	title         = {Shadow of a circular disformal {Kerr} black hole beyond {GR}},
	year          = {2026},
	eprint        = {2601.01767},
	archivePrefix = {arXiv},
	primaryClass  = {gr-qc}
}

@article{WeiWangZhangLiuMann2026,
	author        = {Shao-Wen Wei and Chao-Hui Wang and Yu-Peng Zhang
	and Yu-Xiao Liu and Robert B. Mann},
	title         = {Gravitational equal-area law and critical phenomena
	of cuspy black hole shadow},
	journal       = {Phys. Rev. D},
	year          = {2026},
	note          = {in press},
	doi           = {10.1103/q7b1-4lg6},
	eprint        = {2601.15612},
	archivePrefix = {arXiv},
	primaryClass  = {gr-qc}
}

@article{LiuLiuWuLiu2026,
	author        = {Wentao Liu and Yang Liu and Di Wu and Yu-Xiao Liu},
	title         = {Universal framework for horizon-scale tests of gravity
	with black hole shadows},
	journal       = {Phys. Rev. D},
	volume        = {114},
	pages         = {L021503},
	year          = {2026},
	doi           = {10.1103/66x9-zj16},
	eprint        = {2511.06017},
	archivePrefix = {arXiv},
	primaryClass  = {gr-qc}
}

@article{CunhaHerdeiroRaduRunarsson2015,
	author        = {Pedro V. P. Cunha and Carlos A. R. Herdeiro
	and Eugen Radu and Helgi F. R\'unarsson},
	title         = {Shadows of {Kerr} Black Holes with Scalar Hair},
	journal       = {Phys. Rev. Lett.},
	volume        = {115},
	pages         = {211102},
	year          = {2015},
	doi           = {10.1103/PhysRevLett.115.211102},
	eprint        = {1509.00021},
	archivePrefix = {arXiv},
	primaryClass  = {gr-qc}
}

@article{YounsiZhidenkoRezzollaKonoplyaMizuno2016,
	author        = {Ziri Younsi and Alexander Zhidenko and Luciano Rezzolla
	and Roman Konoplya and Yosuke Mizuno},
	title         = {New method for shadow calculations:
	Application to parametrized axisymmetric black holes},
	journal       = {Phys. Rev. D},
	volume        = {94},
	pages         = {084025},
	year          = {2016},
	doi           = {10.1103/PhysRevD.94.084025},
	eprint        = {1607.05767},
	archivePrefix = {arXiv},
	primaryClass  = {gr-qc}
}

@article{MedeirosPsaltisOzel2020,
	author        = {Lia Medeiros and Dimitrios Psaltis and Feryal {\"O}zel},
	title         = {A Parametric Model for the Shapes of Black Hole Shadows
	in Non-{Kerr} Spacetimes},
	journal       = {Astrophys. J.},
	volume        = {896},
	pages         = {7},
	year          = {2020},
	doi           = {10.3847/1538-4357/ab8bd1},
	eprint        = {1907.12575},
	archivePrefix = {arXiv},
	primaryClass  = {astro-ph.HE}
}

@article{BroderickJohannsenLoebPsaltis2014,
	author        = {Avery E. Broderick and Tim Johannsen
	and Abraham Loeb and Dimitrios Psaltis},
	title         = {Testing the No-Hair Theorem with {Event Horizon Telescope}
	Observations of {Sagittarius A*}},
	journal       = {Astrophys. J.},
	volume        = {784},
	pages         = {7},
	year          = {2014},
	doi           = {10.1088/0004-637X/784/1/7},
	eprint        = {1311.5564},
	archivePrefix = {arXiv},
	primaryClass  = {astro-ph.HE}
}

@article{MizunoYounsiFrommEtAl2018,
	author        = {Yosuke Mizuno and Ziri Younsi and Christian M. Fromm
	and Oliver Porth and Mariafelicia De Laurentis
	and Hector Olivares and Heino Falcke
	and Michael Kramer and Luciano Rezzolla},
	title         = {The Current Ability to Test Theories of Gravity
	with Black Hole Shadows},
	journal       = {Nature Astron.},
	volume        = {2},
	pages         = {585--590},
	year          = {2018},
	doi           = {10.1038/s41550-018-0449-5},
	eprint        = {1804.05812},
	archivePrefix = {arXiv},
	primaryClass  = {astro-ph.GA}
}

@article{UniyalDihingiaMizunoRezzolla2026,
	author        = {Akhil Uniyal and Indu K. Dihingia
	and Yosuke Mizuno and Luciano Rezzolla},
	title         = {The Future Ability to Test Theories of Gravity
	with Black-Hole Shadows},
	journal       = {Nature Astron.},
	volume        = {10},
	pages         = {165--172},
	year          = {2026},
	doi           = {10.1038/s41550-025-02695-4},
	eprint        = {2511.03789},
	archivePrefix = {arXiv},
	primaryClass  = {gr-qc}
}

@article{MengFanLiHanZhang2023,
  author        = {Meng, Kun and Fan, Xi-Long and Li, Song and Han, Wen-Biao and Zhang, Hongsheng},
  title         = {Dynamics of Null Particles and Shadow for General Rotating Black Hole},
  journal       = {J. High Energy Phys.},
  volume        = {2023},
  number        = {11},
  pages         = {141},
  year          = {2023},
  doi           = {10.1007/JHEP11(2023)141},
  eprint        = {2307.08953},
  archivePrefix = {arXiv},
  primaryClass  = {gr-qc}
}

@article{WangXuWei2019,
  author        = {Wang, Hui-Min and Xu, Yu-Meng and Wei, Shao-Wen},
  title         = {Shadows of {Kerr-like} Black Holes in a Modified Gravity Theory},
  journal       = {J. Cosmol. Astropart. Phys.},
  volume        = {2019},
  number        = {03},
  pages         = {046},
  year          = {2019},
  doi           = {10.1088/1475-7516/2019/03/046},
  eprint        = {1810.12767},
  archivePrefix = {arXiv},
  primaryClass  = {gr-qc}
}

@article{LongWangChenJing2019,
  author        = {Long, Fen and Wang, Jieci and Chen, Songbai and Jing, Jiliang},
  title         = {Shadow of a Rotating Squashed {Kaluza--Klein} Black Hole},
  journal       = {J. High Energy Phys.},
  volume        = {2019},
  number        = {10},
  pages         = {269},
  year          = {2019},
  doi           = {10.1007/JHEP10(2019)269},
  eprint        = {1906.04456},
  archivePrefix = {arXiv},
  primaryClass  = {gr-qc}
}

@article{JohannsenPsaltisImages2010,
  author        = {Johannsen, Tim and Psaltis, Dimitrios},
  title         = {Testing the No-Hair Theorem with Observations in the Electromagnetic Spectrum. {II}. Black-Hole Images},
  journal       = {Astrophys. J.},
  volume        = {718},
  pages         = {446--454},
  year          = {2010},
  doi           = {10.1088/0004-637X/718/1/446},
  eprint        = {1005.1931},
  archivePrefix = {arXiv},
  primaryClass  = {astro-ph.HE}
}

@article{WanZhangWeiHouChen2026,
  author        = {Wan, Xi and Zhang, Zhenyu and Wei, Fang-Stars and Hou, Yehui and Chen, Bin},
  title         = {Critical Behavior of Photon Rings in {Kerr--Bertotti--Robinson} Spacetime},
  journal       = {Phys. Rev. D},
  year          = {2026},
  note          = {accepted for publication},
  doi           = {10.1103/fyky-bkbg},
  eprint        = {2603.25049},
  archivePrefix = {arXiv},
  primaryClass  = {gr-qc}
}

@article{PsaltisOzelChanMarrone2015,
  author        = {Psaltis, Dimitrios and {\"O}zel, Feryal and Chan, Chi-Kwan and Marrone, Daniel P.},
  title         = {A General Relativistic Null Hypothesis Test with {Event Horizon Telescope} Observations of the Black Hole Shadow in {Sgr A*}},
  journal       = {Astrophys. J.},
  volume        = {814},
  number        = {2},
  pages         = {115},
  year          = {2015},
  doi           = {10.1088/0004-637X/814/2/115},
  eprint        = {1411.1454},
  archivePrefix = {arXiv},
  primaryClass  = {astro-ph.HE}
}

@article{EHTM87Metric2020,
  author        = {{Event Horizon Telescope Collaboration}},
  title         = {Gravitational Test Beyond the First Post-{Newtonian} Order with the Shadow of the {M87} Black Hole},
  journal       = {Phys. Rev. Lett.},
  volume        = {125},
  pages         = {141104},
  year          = {2020},
  doi           = {10.1103/PhysRevLett.125.141104},
  eprint        = {2010.01055},
  archivePrefix = {arXiv},
  primaryClass  = {gr-qc}
}

@article{EHTSgrAMetric2022,
  author        = {{Event Horizon Telescope Collaboration}},
  title         = {First {Sagittarius A*} {Event Horizon Telescope} Results. {VI}. Testing the Black Hole Metric},
  journal       = {Astrophys. J. Lett.},
  volume        = {930},
  number        = {2},
  pages         = {L17},
  year          = {2022},
  doi           = {10.3847/2041-8213/ac6756},
  eprint        = {2311.09484},
  archivePrefix = {arXiv},
  primaryClass  = {astro-ph.HE}
}

@article{ChenJingQianWang2023,
  author        = {Chen, Songbai and Jing, Jiliang and Qian, Wei-Liang and Wang, Bin},
  title         = {Black Hole Images: A Review},
  journal       = {Sci. China Phys. Mech. Astron.},
  volume        = {66},
  number        = {6},
  pages         = {260401},
  year          = {2023},
  doi           = {10.1007/s11433-022-2059-5},
  eprint        = {2301.00113},
  archivePrefix = {arXiv},
  primaryClass  = {astro-ph.HE}
}

@article{LiMirzaevAbdujabbarovMalafarinaAhmedovHan2022,
  author        = {Li, Song and Mirzaev, Temurbek and Abdujabbarov, Ahmadjon A. and Malafarina, Daniele and Ahmedov, Bobomurat and Han, Wen-Biao},
  title         = {Constraining the Deformation of a Rotating Black Hole Mimicker from Its Shadow},
  journal       = {Phys. Rev. D},
  volume        = {106},
  pages         = {084041},
  year          = {2022},
  doi           = {10.1103/PhysRevD.106.084041},
  eprint        = {2207.10933},
  archivePrefix = {arXiv},
  primaryClass  = {gr-qc}
}

@article{MengKuangTang2022,
  author        = {Meng, Yuan and Kuang, Xiao-Mei and Tang, Zi-Yu},
  title         = {Photon Regions, Shadow Observables, and Constraints from {M87*} of a Charged Rotating Black Hole},
  journal       = {Phys. Rev. D},
  volume        = {106},
  pages         = {064006},
  year          = {2022},
  doi           = {10.1103/PhysRevD.106.064006},
  eprint        = {2204.00897},
  archivePrefix = {arXiv},
  primaryClass  = {gr-qc}
}

@article{WangChenJing2021,
  author        = {Wang, Mingzhi and Chen, Songbai and Jing, Jiliang},
  title         = {{Kerr} Black Hole Shadows in {Melvin} Magnetic Field with Stable Photon Orbits},
  journal       = {Phys. Rev. D},
  volume        = {104},
  pages         = {084021},
  year          = {2021},
  doi           = {10.1103/PhysRevD.104.084021},
  eprint        = {2104.12304},
  archivePrefix = {arXiv},
  primaryClass  = {gr-qc}
}

@article{GanWangWuYang2021,
  author        = {Gan, Qingyu and Wang, Peng and Wu, Houwen and Yang, Haitang},
  title         = {Photon Ring and Observational Appearance of a Hairy Black Hole},
  journal       = {Phys. Rev. D},
  volume        = {104},
  pages         = {044049},
  year          = {2021},
  doi           = {10.1103/PhysRevD.104.044049},
  eprint        = {2105.11770},
  archivePrefix = {arXiv},
  primaryClass  = {gr-qc}
}

@article{WuWei2023,
  author        = {Wu, Shan-Ping and Wei, Shao-Wen},
  title         = {Topology of Light Rings for Extremal and Nonextremal {Kerr--Newman--Taub--NUT} Black Holes without $\mathbb{Z}_2$ Symmetry},
  journal       = {Phys. Rev. D},
  volume        = {108},
  pages         = {104041},
  year          = {2023},
  doi           = {10.1103/PhysRevD.108.104041},
  eprint        = {2307.14003},
  archivePrefix = {arXiv},
  primaryClass  = {gr-qc}
}

@article{ZhangJiang2021,
  author        = {Zhang, Ming and Jiang, Jie},
  title         = {{NUT} Charges and Black Hole Shadows},
  journal       = {Phys. Lett. B},
  volume        = {816},
  pages         = {136213},
  year          = {2021},
  doi           = {10.1016/j.physletb.2021.136213},
  eprint        = {2103.11416},
  archivePrefix = {arXiv},
  primaryClass  = {gr-qc}
}

@article{GrenzebachPerlickLammerzahl2014,
  author        = {Grenzebach, Arne and Perlick, Volker and L{\"a}mmerzahl, Claus},
  title         = {Photon Regions and Shadows of {Kerr--Newman--NUT} Black Holes with a Cosmological Constant},
  journal       = {Phys. Rev. D},
  volume        = {89},
  pages         = {124004},
  year          = {2014},
  doi           = {10.1103/PhysRevD.89.124004},
  eprint        = {1403.5234},
  archivePrefix = {arXiv},
  primaryClass  = {gr-qc}
}

@article{HeLiYangZeng2025,
	author        = {Ke-Jian He and Guo-Ping Li and Chen-Yu Yang and Xiao-Xiong Zeng},
	title         = {Observational features of the rotating {Bardeen} black hole
	surrounded by perfect fluid dark matter},
	journal       = {Eur. Phys. J. C},
	volume        = {85},
	pages         = {662},
	year          = {2025},
	doi           = {10.1140/epjc/s10052-025-14391-8},
	eprint        = {2411.11680},
	archivePrefix = {arXiv},
	primaryClass  = {astro-ph.HE}
}

@article{ZengYangAslamSaleem2026,
	author        = {Xiao-Xiong Zeng and Chen-Yu Yang
	and Muhammad Israr Aslam and Rabia Saleem},
	title         = {Probing {Horndeski} gravity via {Kerr} black hole:
	Insights from thin accretion disks and shadows
	with {EHT} observations},
	journal       = {J. High Energy Astrophys.},
	volume        = {51},
	pages         = {100540},
	year          = {2026},
	doi           = {10.1016/j.jheap.2025.100540},
	eprint        = {2509.05803},
	archivePrefix = {arXiv},
	primaryClass  = {gr-qc}
}

@article{HeYeZengLiXu2025,
	author        = {Ke-Jian He and Huan Ye and Xiao-Xiong Zeng
	and Li-Fang Li and Peng Xu},
	title         = {Shadow and accretion disk images of the rotation
	loop quantum black bounce},
	journal       = {Chin. Phys. C},
	volume        = {49},
	pages         = {125103},
	year          = {2025},
	doi           = {10.1088/1674-1137/adf4a2},
	eprint        = {2502.08388},
	archivePrefix = {arXiv},
	primaryClass  = {gr-qc}
}

@article{HouZhangYanGuoChen2022,
  author        = {Hou, Yehui and Zhang, Zhenyu and Yan, Haopeng and Guo, Minyong and Chen, Bin},
  title         = {Image of a {Kerr--Melvin} Black Hole with a Thin Accretion Disk},
  journal       = {Phys. Rev. D},
  volume        = {106},
  number        = {6},
  pages         = {064058},
  year          = {2022},
  doi           = {10.1103/PhysRevD.106.064058},
  eprint        = {2206.13744},
  archivePrefix = {arXiv},
  primaryClass  = {gr-qc}
}

@article{ZhangHouGuoChen2024,
  author        = {Zhang, Zhenyu and Hou, Yehui and Guo, Minyong and Chen, Bin},
  title         = {Imaging Thick Accretion Disks and Jets Surrounding Black Holes},
  journal       = {J. Cosmol. Astropart. Phys.},
  volume        = {2024},
  number        = {05},
  pages         = {032},
  year          = {2024},
  doi           = {10.1088/1475-7516/2024/05/032},
  eprint        = {2401.14794},
  archivePrefix = {arXiv},
  primaryClass  = {astro-ph.HE}
}

@article{WangWangZhangGuo2024,
  author        = {Wang, Xiangyu and Wang, Xiaobao and Zhang, Hai-Qing and Guo, Minyong},
  title         = {Is a Photon Ring Invariably a Closed Structure?},
  journal       = {Eur. Phys. J. C},
  volume        = {84},
  pages         = {1168},
  year          = {2024},
  doi           = {10.1140/epjc/s10052-024-13527-6},
  eprint        = {2405.05011},
  archivePrefix = {arXiv},
  primaryClass  = {gr-qc}
}

@article{Gui:2024owz,
	author        = {Jin-Yu Gui and Ke-Jian He and Huan Ye and Xiao-Xiong Zeng},
	title         = {Holographic {Einstein} ring of deformed
	{AdS--Schwarzschild} black holes},
	journal       = {Front. Phys.},
	volume        = {20},
	number        = {2},
	pages         = {025202},
	year          = {2025},
	doi           = {10.15302/frontphys.2025.025202},
	eprint        = {2407.09069},
	archivePrefix = {arXiv},
	primaryClass  = {gr-qc}
}

@article{Zeng:2024ptv,
	author        = {Xiao-Xiong Zeng and Li-Fang Li and Pan Li
	and Bo Liang and Peng Xu},
	title         = {Holographic images of a charged black hole in
	{Lorentz} symmetry breaking massive gravity},
	journal       = {Sci. China Phys. Mech. Astron.},
	volume        = {68},
	number        = {2},
	pages         = {220412},
	year          = {2025},
	doi           = {10.1007/s11433-024-2526-4},
	eprint        = {2411.12528},
	archivePrefix = {arXiv},
	primaryClass  = {gr-qc}
}

@article{Huang:2024bar,
	author = "Huang, Jiewei and Zheng, Liheng and Guo, Minyong and Chen, Bin",
	title = "{Coport: a new public code for polarized radiative transfer in a covariant framework}",
	eprint = "2407.10431",
	archivePrefix = "arXiv",
	primaryClass = "astro-ph.HE",
	doi = "10.1088/1475-7516/2024/11/054",
	journal = "JCAP",
	volume = "11",
	pages = "054",
	year = "2024"
}

@article{Li:2021zct,
	author = "Li, Peng-Cheng and Lee, Tsai-Chen and Guo, Minyong and Chen, Bin",
	title = "{Correspondence of eikonal quasinormal modes and unstable fundamental photon orbits for a Kerr-Newman black hole}",
	eprint = "2105.14268",
	archivePrefix = "arXiv",
	primaryClass = "gr-qc",
	doi = "10.1103/PhysRevD.104.084044",
	journal = "Phys. Rev. D",
	volume = "104",
	number = "8",
	pages = "084044",
	year = "2021"
}
	
\end{document}